\documentclass[journal,onecolumn,draftcls,12pt]{IEEEtran}
\usepackage{subfigure}
\usepackage{comment}
\usepackage{algorithm2e,algorithmic}
\usepackage{cite}
\usepackage{booktabs}
\usepackage{bm,bbm}
\usepackage{graphicx,psfrag,float}
\usepackage{siunitx}
\usepackage{amsmath, amsthm, amsfonts}    
\usepackage{bbm}
\usepackage{booktabs}
\usepackage{multirow}

\makeatletter
\def\endthebibliography{%
    \def\@noitemerr{\@latex@warning{Empty `thebibliography' environment}}%
    \endlist
}
\renewcommand{\@algocf@capt@plain}{above}
\makeatother

\begin{document}
\title{Comparing Samples from the $\mathcal{G}^0$ Distribution using a Geodesic Distance}
\author{ Alejandro C.\ Frery, Juliana Gambini

\thanks{Alejandro C.\ Frery is with the Laborat\'orio de Computa\c c\~ao Cient\'ifica e An\'alise Num\'erica,
Universidade Federal de Alagoas, Av.\ Lourival Melo Mota, s/n, 57072-900, Macei\'o -- AL, Brazil, 
 \texttt{acfrery@gmail.com}}

\thanks{Juliana Gambini is with the Depto. de Ingenier\'{\i}a Inform\'atica, Instituto Tecnol\'ogico de Buenos Aires, Av.\ Madero 399, C1106ACD Buenos Aires,  Argentina and with Depto. de Ingenier\'{\i}a en Computaci\'on, Universidad Nacional de Tres de Febrero, Pcia.\ de Buenos Aires, Argentina, \texttt{juliana.gambini@gmail.com}.}
}

\maketitle

\begin{abstract} 
The $\mathcal{G}^0$ distribution is widely used for monopolarized SAR image modeling because it can characterize regions with different degree of texture accurately. It is indexed by three parameters: the number of looks (which can be estimated for the whole image), a scale parameter and a texture parameter.
This paper presents a new proposal for comparing samples from the $\mathcal{G}^0$ distribution using a Geodesic Distance (GD) as a measure of dissimilarity between models.
The objective is quantifying the difference between pairs of samples from SAR data using both local parameters (scale and texture) of the $\mathcal{G}^0$ distribution.
We propose three tests based on the GD which combine the tests presented in~\cite{GeodesicDistanceGI0JSTARS}, and we estimate their probability distributions using permutation methods. 
\end{abstract}

\textbf{Keywords}: Geodesic Distance, Dissimilarity Measure, $\mathcal{G}^0$ Distribution

\section{Introduction}

Automatic detection of differences between samples from SAR (\textit{Synthetic Aperture Radar}) images is both challenging and necessary.
It has important applications in, among others,
urban planning~\cite{6352307}, 
disaster management~\cite{6235981}, 
emergency response~\cite{7502166}, 
environmental monitoring, and ecology~\cite{Hill2005}.  
The main idea is developing methods for automatic discrimination of regions with different levels of texture and/or roughness.
As in~\cite{Gambiniijrs,GambiniSC08}, we adopt the $\mathcal{G}^0$ distribution as model for the data.

The $\mathcal{G}^0$ distribution is widely used for monopolarized SAR image modeling because it can characterize different regions accurately.
It is indexed by three parameters: the number of looks $L$ (which can be estimated for the whole image), a scale parameter $\gamma$, and a texture parameter $\alpha$. 
The last two are local parameters and relate directly to the target.

Nacimento et al.~\cite{5208318} obtained test statistics based on Information Theory to assess the null hypothesis that two samples were produced by the same $\mathcal{G}^0$ law, provided the same number of looks is known.
The approach consisted of first computing $h$-$\phi$ divergences between the models, indexing their symmetrized versions with maximum likelihood estimates and scaling appropriately to obtain test statistics.
These tests, under mild regularity conditions, follow asymptotically $\chi^2$ laws.
These divergences and associated test statistics were successfully applied to region discrimination~\cite{ClassificationPolSARSegmentsMinimizationWishartDistances}, 
segmentation~\cite{SARSegmentationLevelSetGA0}, and
parameter estimation~\cite{gambini2015}.

Two issues make their use somewhat difficult, though, namely 
(i)~they require the numerical integration of expressions that, more often than not, involve special functions, and
(ii)~the choice of the particular test statistic might be considered arbitrary (different choices of the functions $h$ and $\phi$ lead, among infinitely many others, to the Kullback-Leibler, Hellinger, Bhattacharya, Triangular, Harmonic, Jensen-Shannon, and R\'enyi of order $\beta$ divergences).
The Geodesic Distance solves the second difficulty, as it is unique, and gives a partial solution to the first one.

The Geodesic Distance can be used to measure the difference between two parametric distributions.
It was presented by Rao~\cite{raey1945,raey1992}, and since then it has been studied by several authors~\cite{ISI:000302346300002,ISI:A1995RC74900013,AtkinsonMitchell1981}.
In Ref.~\cite{ISI:000301842800003,Statisticalhypothesistestforrobustclassification2015}, it is used as measure of contrast between samples by means of statistical tests presented in~\cite{OntheApplicationsofDivergenceTypeMeasuresinTestingStatisticalHypothesessalicru,ISI:A1995RC74900013,5208318}, where the authors demonstrated that its distribution is $\chi^2_1$.

To the best of the authors' knowledge, there is no closed expression for the geodesic distance between two $\mathcal{G}^0$ models with both $\alpha$ and $\gamma$ unknown, given $L$.
In this work, we analyze several statistical hypothesis tests depending on both parameters to discriminate two samples from $\mathcal{G}^0$ models with both parameters unknown.
We use permutation methods to estimate the distribution of such tests statistics since no explicit results are available.

The paper unfolds as follows.
Section~\ref{sec_SAR} recalls properties of the $\mathcal{G}^0$ model, including parameter estimation by maximum likelihood.
Section~\ref{geodesicdistances} presents the expressions for the GD with one parameter known.
Section~\ref{Sec:OneParameter} analyzes the behavior of the test statistics based on a known parameter.
In Section~\ref{Sec:TwoParameter} we study the more realistic situation of estimating both scale and texture, while assuming known the number of looks.
Finally, in Section~\ref{conclu} we present conclusions and outline future work.

\section{SAR Imagery and the $\mathcal G^0$ Model}
\label{sec_SAR}

Under the multiplicative model, the return in monopolarized SAR images can be modeled as the product of two independent random variables, one corresponding to the backscatter $X$ and other to the
speckle noise $Y$.
In this manner, $Z=X  Y$ models the return $Z$ in each pixel.
For monopolarized data, speckle $Y$ is modeled as a $\Gamma$ distributed random variable with unitary mean and shape parameter $L$, the number of looks.
A good choice for the backscater distribution $X$ is the  reciprocal of Gamma $\Gamma ^{-1}( \alpha ,\gamma ) $ law that gives rise to the $\mathcal{G}^{0}$ distribution for the return $Z$~\cite{Frery97}.
The mathematical tractability and descriptive power of the $\mathcal{G}^{0}$ distribution make it an attractive choice for SAR data modeling~\cite{QuartulliDatcu:04}.
The probability density function for intensity data under the $\mathcal G^0(\alpha, \gamma,L)$ distribution is:
\begin{equation}
f_{\mathcal{G}^{0}}( z) =\frac{L^{L}\Gamma ( L-\alpha
    ) }{\gamma ^{\alpha }\Gamma ( -\alpha ) \Gamma (
    L) }
\frac{z^{L-1}}{( \gamma +zL) ^{L-\alpha }},%
\label{ec_dens_gI0}
\end{equation}
where $-\alpha,\gamma ,z>0$ and $L\geq 1$.
If $\alpha\to-\infty$, the $\mathcal{G}^0$ distribution becomes an exponential law.
The $r$-order moments are given by
\begin{equation}
E(Z^r) =\Big(\frac{\gamma}{L}\Big)^r\frac{\Gamma ( -\alpha-r )}{ \Gamma (-\alpha) }  
\frac{\Gamma (L+r )}{\Gamma (L)}.
\label{moments_gI0}
\end{equation}

To simplify calculation and with the intention of obtaining comparable results, in most experiments, we deal with a restricted case which assumes $E(Z)=1$.

Using that 
$\Gamma(L + 1) = L \Gamma(L)$ and that $\Gamma(-\alpha) = (-\alpha-1) \Gamma(-\alpha-1)$ in~\eqref{moments_gI0}, assuming $L=1$ and imposing $E(Z) =1$ we find the following relation between $\alpha$ and $\gamma$:
\begin{equation*}
\gamma^* = -\alpha-1.
\end{equation*}    
Then, the random variable $Z$ with $\mathcal{G}_I^0(\alpha, \gamma^*, 1)$ distribution has unitary mean.
This allows us to simplify the calculations and to obtain results which do not depend on image brightness.

One of the essential features of the $\mathcal{G}^0$ distribution is the ability to interpret its parameters. 
The $\alpha$ parameter is a texture parameter, which is related to the roughness or number of elementary backscatterers of the target. 
Values close to zero (typically above $-3$) suggest extremely textured targets, as urban zones. 
As the value decreases, it indicates regions with moderate texture (usually $\alpha \in [-6,-3]$), as forest zones.
Textureless targets, e.g. pasture, usually produce $\alpha\in(-\infty,-6)$. 

The $\gamma$ parameter of the $\mathcal{G}^0$ distribution is a scale parameter, that is, if $W\sim\mathcal{G}^0(\alpha, \gamma, L)$, then 
${\gamma}^{-1}W \sim \mathcal{G}^0(\alpha, 1, L)$.

\begin{figure}[hbt]
\centering
\subfigure[Densities in linear scale\label{fig:denslinear}]{\includegraphics[width=.45\linewidth]{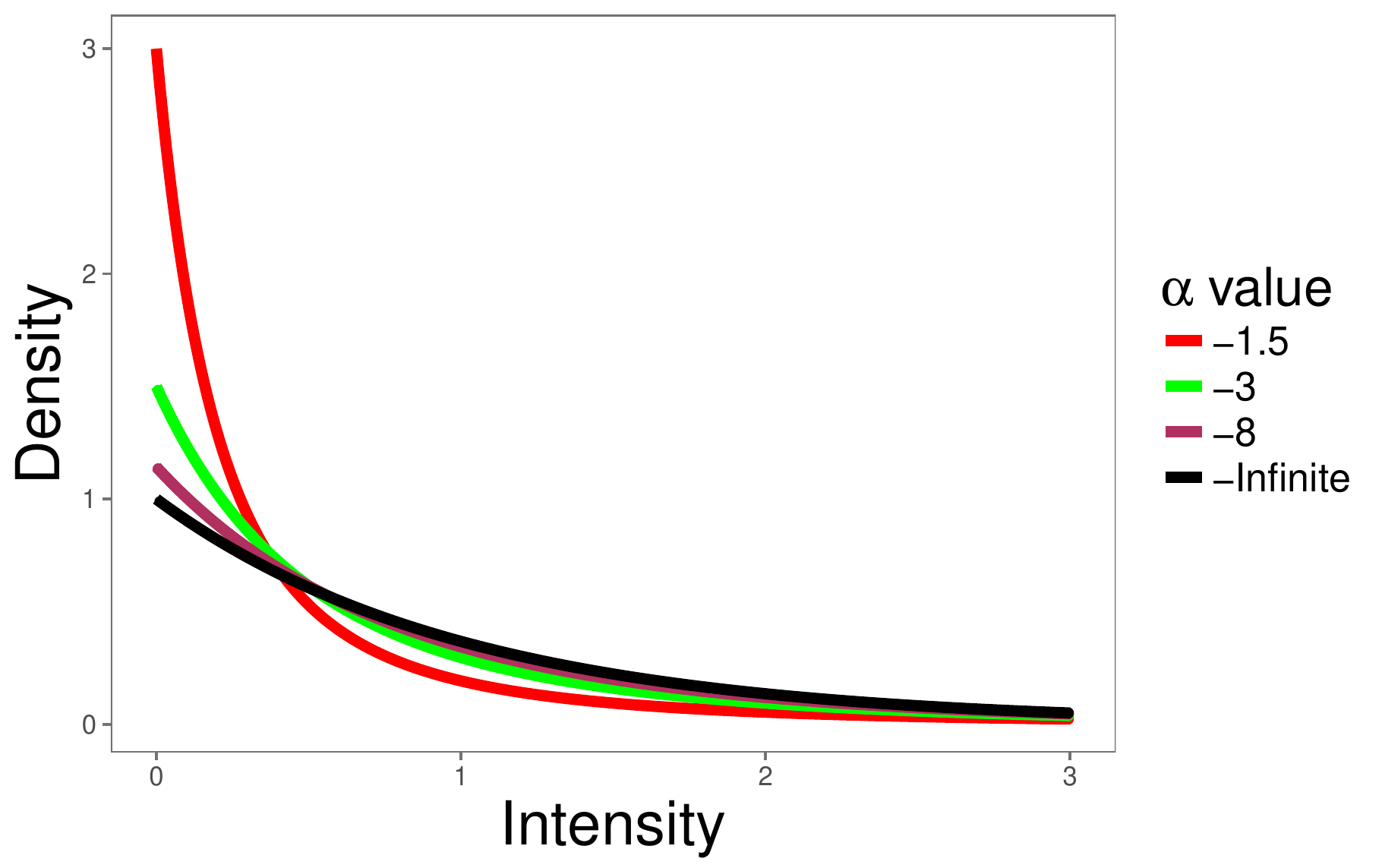}}
\subfigure[Densities in semilogarithmic scale\label{fig:denssemilog}]{\includegraphics[width=.45\linewidth]{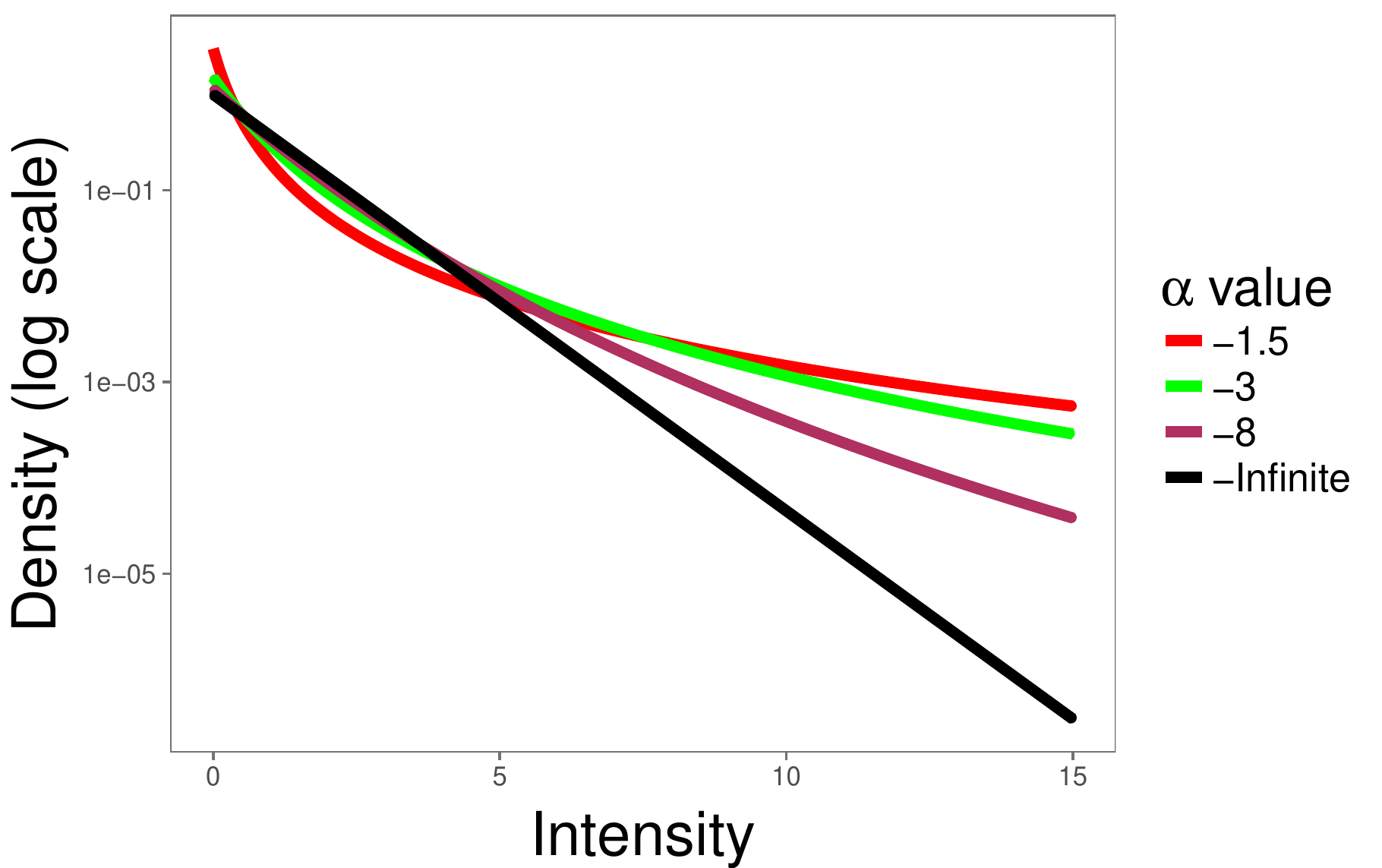}}
\caption{Densities of $\mathcal G^0(\alpha,\gamma^*, 1)$ distributions for $\alpha\in\{-\infty, -8, -3,-1.5\}$ (black, maroon, green, red, respectively).}\label{fig:densities}
\end{figure}

Fig.~\ref{fig:densities} shows the densities of $\mathcal G^0(\alpha,\gamma^*, 1)$ distributions for $\alpha\in\{-\infty, -8, -3,-1.5\}$ (black, maroon, green, red, respectively) in linear (Fig.~\ref{fig:denslinear}) and semi-logarithmic (Fig.~\ref{fig:denssemilog}) scales.

The difference between these densities becomes more apparent in semi-logarithmic scale, where the limiting distribution (for $\alpha\to-\infty$ ) appears as a straight line.
The larger $\alpha$ is, the more prone the random variable to produce extreme values is.

Given the sample  $\bm{z}=(z_1,\dots, z_n)$ of independent and identically distributed random variables with common distribution $\mathcal{G}^0(\alpha,\gamma,L)$ with $(\alpha,\gamma) \in \Theta$, $\Theta = \mathbbm{R}_- \times \mathbbm{R}_+ $, a maximum likelihood estimator of $(\alpha,\gamma)$ satisfies 
\begin{equation*}
(\widehat{\alpha},\widehat{\gamma}) = \arg\max_{ (\alpha,\gamma) \in \Theta}\mathcal{L}(\alpha,\gamma,L,\bm{z}),
\end{equation*}
where $\mathcal L$ is the likelihood function under the  $\mathcal{G}^0(\alpha,\gamma,L)$ distribution.
This leads to $\widehat{\alpha}$ and $\widehat{\gamma}$ such that

\begin{eqnarray}
n[\Psi^0(-\widehat{\alpha})-\Psi^0(L-\widehat{\alpha})]+\sum_{i=1}^n \ln\frac{\widehat{\gamma}+Lz_i^2}{\widehat{\gamma}}=0\\
\frac{n\widehat{\alpha}}{\widehat{\gamma}}+(L-\widehat{\alpha})\sum_{i=1}^n(\widehat{\gamma}+Lz_i)^{-1}=0,
\end{eqnarray}
where $\Psi^0(t) = {d\ln\Gamma(t)}/{dt}$ is the digamma function. In many cases no explicit solution for this system is available and numerical methods have to be used. 
In this work, we applied the BFGS~\cite{Broyden65} optimization algorithm.

\section{Geodesic Distance between $\mathcal{G}^0$ Models}
\label{geodesicdistances}

Naranjo-Torres et al.~\cite{GeodesicDistanceGI0JSTARS} obtained two cases of geodesic distances between $\mathcal{G}^0$ distributions with a known number of looks:
the cases where either the texture or the scale is known.
These are given, respectively by
\begin{align}
s(\alpha_{1},\alpha_{2})&= 
\left|\int_{\alpha_{1}}^{\alpha_{2}}\sqrt{
    \sum _{n=1}^L {(-\alpha +n-1)^{-2}}
}
\,d\alpha\right|, \text{ and by} \label{DG_gi0_c1}\\ 
s(\gamma _{1},\gamma _{2})&=\left|\sqrt{\frac{-\alpha L}{-\alpha +L+1}} \ln \frac{\gamma_1}{\gamma_2}\right|.
\label{DG_gi0_c2}
\end{align}

The first equation can be solved explicitly for $L=\left\{1,2\right\}$:
\begin{align*}
s(\alpha_{1},\alpha_{2})\Big|_{L=1} &=\left|\ln \frac{\alpha_1}{\alpha_2}\right|,\text{ and}\label{DG_gi0_cL1}\\
s(\alpha_{1},\alpha_{2})\Big|_{L=2} &=
\left|\ln \frac{\alpha_1^2 (\alpha_2 -1)^2  (\alpha_2 R_2 -1)((\alpha_1 -1) R_1 +1)}{\alpha_2^2(\alpha_1 -1)^2  (\alpha_1 R_1 -1) ((\alpha_2 -1) R_2 +1)} + \sqrt{2} \ln \frac{ 1+\alpha_2 (R_2 -2)- \alpha_2^2 R_2}{1+\alpha_1 (R_1 -2)- \alpha_1^2 R_1} \right|,
\end{align*}
where $R_1=R\left(\alpha_1\right)$ and $R_2=R\left(\alpha_2\right)$ are given by
\[
R\left(\alpha\right)=\sqrt{\frac{4 \alpha^2-4 \alpha+2}{(\alpha-1)^2 \alpha^2}}.
\]

Notice that $s(\gamma _{1},\gamma _{2})$ depends on the texture $\alpha$, while $s(\alpha_{1},\alpha_{2})$ is independent of the scale $\gamma$.
Both~\eqref{DG_gi0_c1} and~\eqref{DG_gi0_c2} depend on the number of looks $L$.

To the best of the authors' knowledge, there is no closed expression for the geodesic distance between two $\mathcal{G}^0$ models with both $\alpha$ and $\gamma$ different, given $L$ known.

Both distances can be turned into test statistics (see~\cite{ISI:000301842800003,Statisticalhypothesistestforrobustclassification2015}) by indexing with maximum likelihood estimators based on samples of sizes $m$ and $n$, and then rescaling:
$$
T = \frac{mn}{m+n}\widehat{s}^2.
$$
We will denote
\begin{align}
T_\alpha &= \frac{mn}{m+n} \big(s(\widehat\alpha_{1},\widehat\alpha_{2})\big)^2, \text{ and} \label{eq:Ta}\\
T_\gamma &= \frac{mn}{m+n} \big(s(\widehat\gamma_{1},\widehat\gamma_{2})\big)^2. \label{eq:Tg}
\end{align}
Under the null hypothesis of equal parameters, when $m,n\to\infty$ proportionally, both $T_\alpha$ and $T_\gamma$ follow a $\chi^2_1$ distribution, so it is possible to compute the $p$-value of two samples under $H_0$ and either reject or not this hypothesis~\cite{ISI:A1995RC74900013}.

Section~\ref{Sec:OneParameter} presents an analysis of the behavior of these test statistics $T_\alpha$ and $T_\gamma$.
Section~\ref{Sec:TwoParameter} studies ways of combining them to produce a two-parameter test.

\section{Analysis of One-parameter tests}\label{Sec:OneParameter}

In this Section, we analyze the finite sample size behavior of the test statistics defined in~\eqref{eq:Ta} and~\eqref{eq:Tg} using Monte Carlo experiments.
We obtained the samples following the guidelines presented in Ref.~\cite{SamplingfromtheGI0Distribution2018}.

The parameter space for the first experiment was
$\alpha=-1.5$ and the same sample size $n\in\{50, 100, 150, \dots, 1000\}$ for
$\gamma=1$ and $L=1$.
We obtained five thousand independent replications for each sample size, and maximized the following reduced log-likelihood function:
\begin{equation}
\ell(\alpha;\gamma,L,\bm z) = n[\log\Gamma(L-\alpha) - \alpha\log\gamma -\log \Gamma(-\alpha)]  
 +  \alpha \sum_{i=1}^n \log(\gamma+L z_i).
\label{eq:LogVera_gknown}
\end{equation}
We produced two independent samples in each replication in order to compute a distance from the respective estimated models.

\begin{figure}[hbt]
    \centering
    \subfigure[Sample densities of $\widehat{\alpha}$.\label{fig:DensMLAlpha_gknown}]{\includegraphics[width=.49\linewidth]{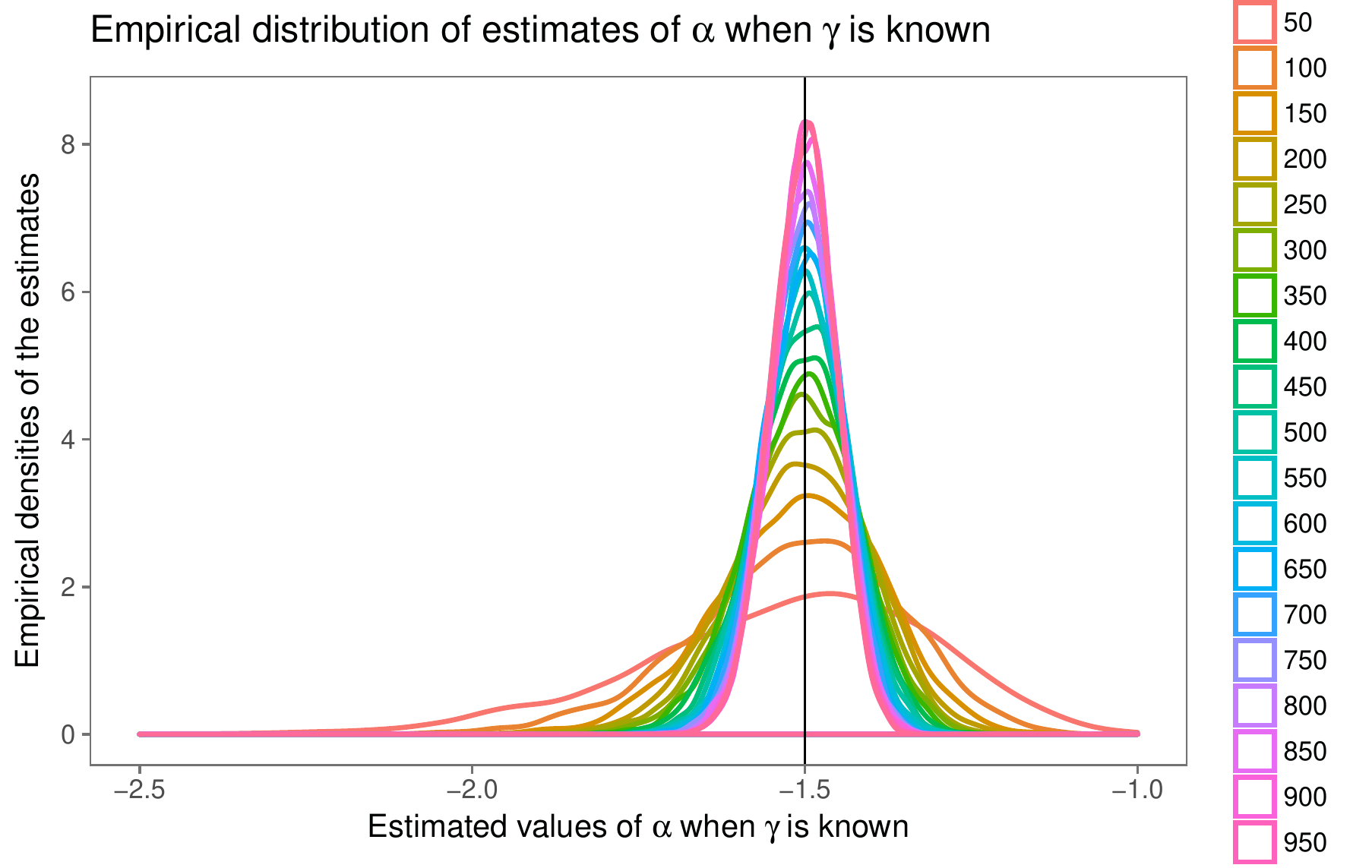}}
    \subfigure[Sample densities of $\widehat{\gamma}$.\label{fig:DensMLGamma_aknown}]{\includegraphics[width=.49\linewidth]{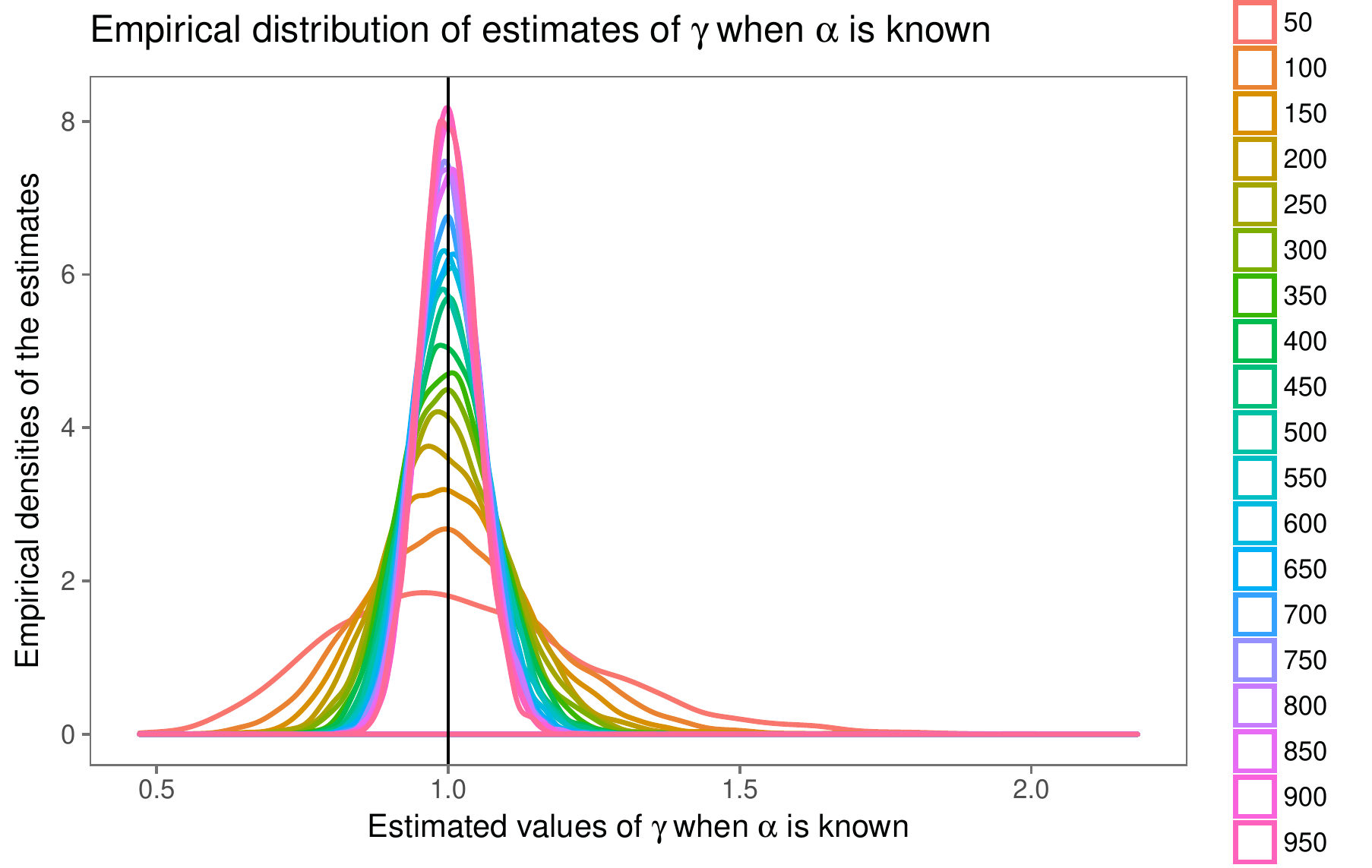}}
    \caption{Estimated densities of maximum likelihood estimators of $\alpha$ and $\gamma$, when only one parameter is unknown and $L=1$.}
    \label{fig:DensMLAlphaGamma_OneUnknown}
\end{figure}

Fig~\ref{fig:DensMLAlpha_gknown} presents the sample densities of $\widehat{\alpha}$ for $\gamma=1$ and $L=1$.
They are all centered around the true value $\alpha=-1.5$ and, as expected, the larger the sample size $n$ is, the smaller the variability is.
Small values of $n$ yield more asymmetric densities than their larger counterpart.
The parameter space and number of replications for the second experiment were the same, but the reduced log-likelihood to be maximized was
\begin{equation}
\ell(\gamma;\alpha,L,\bm z) = -n\alpha\log\gamma + (\alpha-L) \sum_{i=1}^n \log(\gamma+L z_i).
\label{eq:LogVerg_aknown}
\end{equation}

Similar conclusions can be drawn from the sample densities of the maximum likelihood estimators of $\gamma$, when $\alpha$ and $L$ are known; cf.\ Fig.~\ref{fig:DensMLGamma_aknown}.

The behavior shown in Fig.~\ref{fig:DensMLAlphaGamma_OneUnknown} is consistent across other values of $\alpha$ and $\gamma$.
Both~\eqref{eq:LogVera_gknown} and~\eqref{eq:LogVerg_aknown}, as well as the two-parameter reduced log-likelihood function presented below were optimized using the \texttt{maxLik} routine~\cite{maxLik} available in \texttt R~\cite{Rmanual}.

\begin{figure}[hbt]
    \centering
    \subfigure[Proportion of estimates of $\alpha$ whose error is larger than $0.10$, $0.11$, $0.12$, $0.13$.\label{fig:ErrorEstimationAlpha_gknown}]{\includegraphics[width=.49\linewidth]{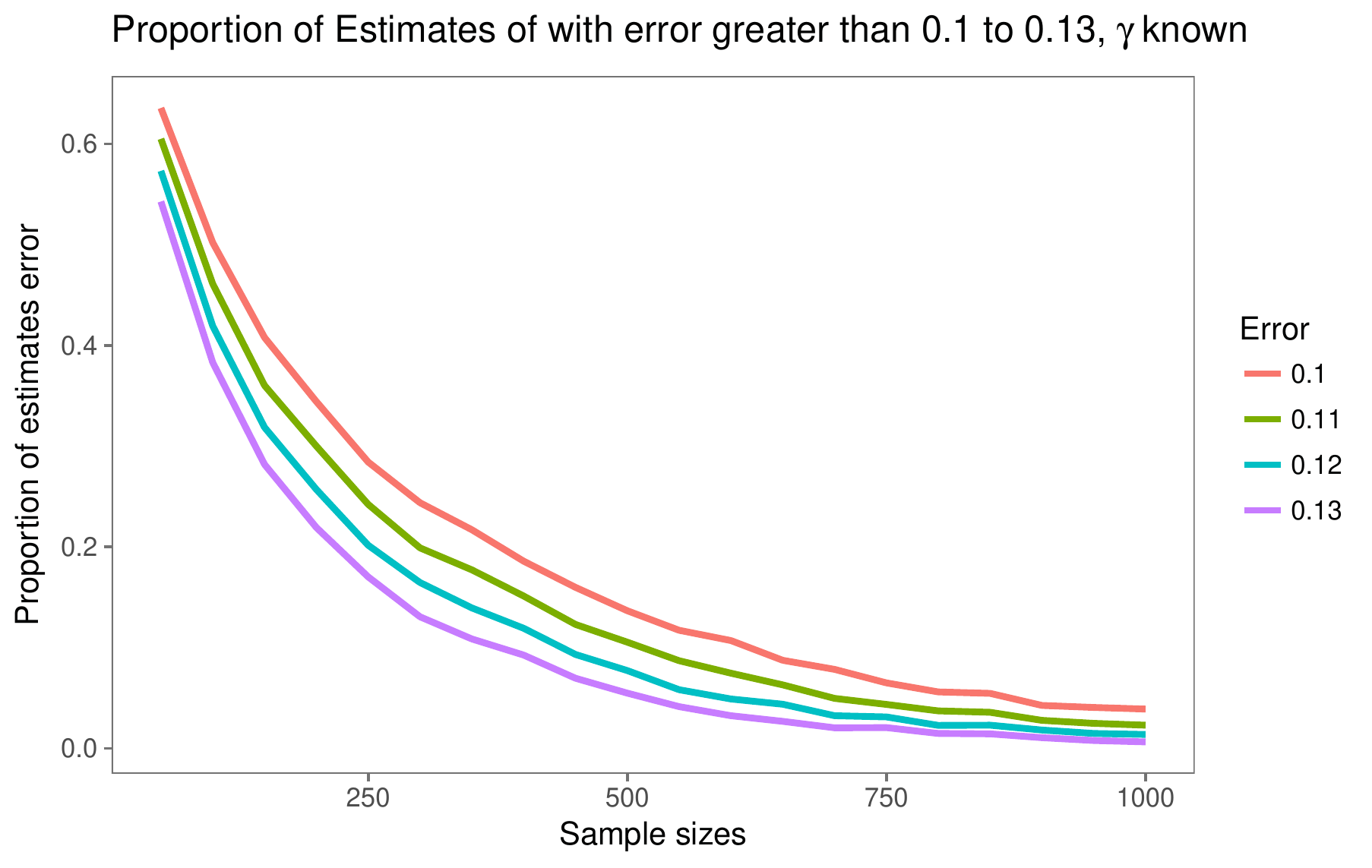}}
    \subfigure[Proportion of estimates of $\gamma$ whose error is larger than $0.10$, $0.11$, $0.12$, $0.13$. \label{fig:ErrorEstimationGamma_aknown}]{\includegraphics[width=.49\linewidth]{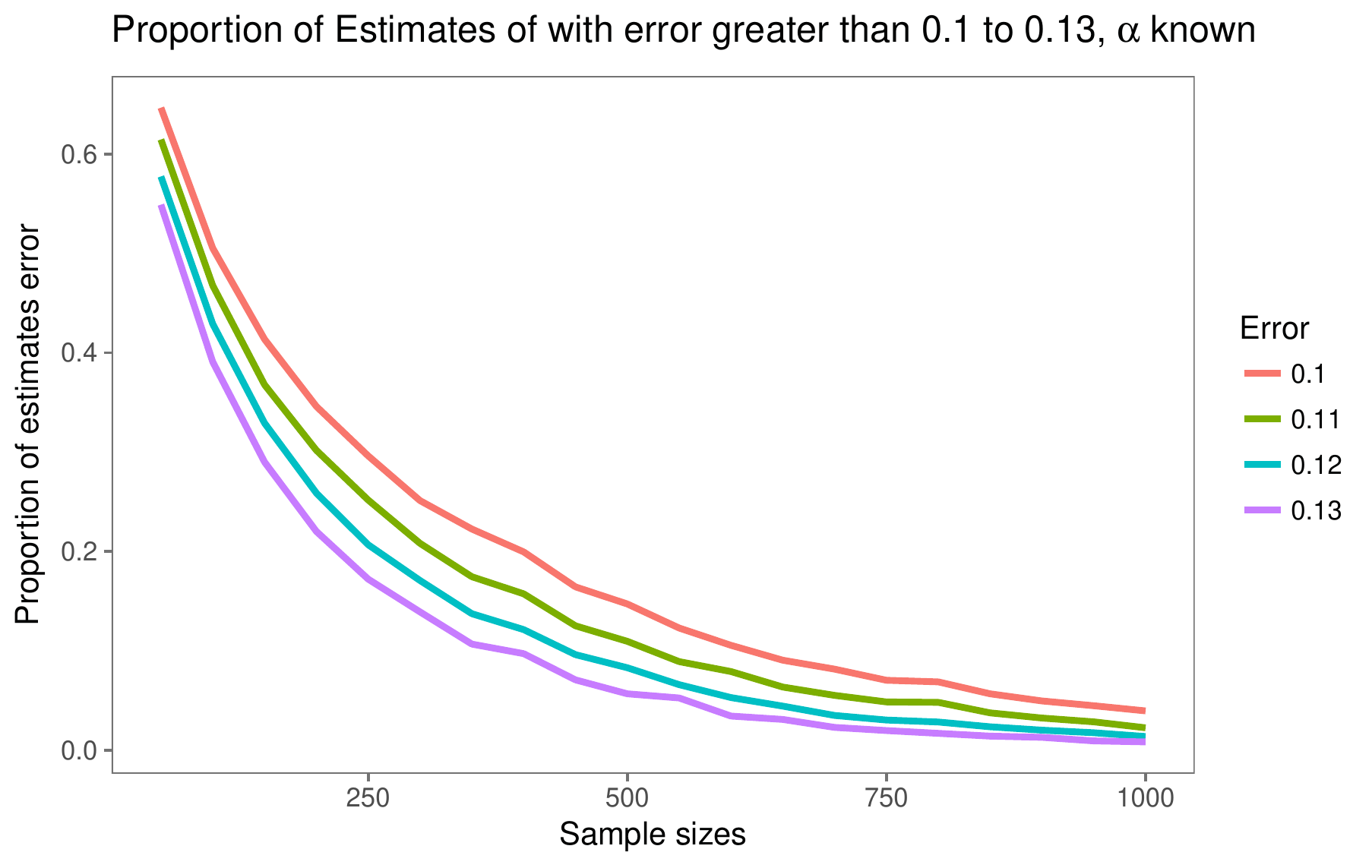}}
    \caption{Proportion of $\alpha$ and $\gamma$ test statistics with errors larger than $0.10$, $0.11$, $0.12$, $0.13$.}
\end{figure}

Figures~\ref{fig:ErrorEstimationAlpha_gknown} and~\ref{fig:ErrorEstimationGamma_aknown} show the proportion of estimates whose error is larger than $0.10$, $0.11$, $0.12$, $0.13$, when the $\gamma$ parameter is known and when the $\alpha$ parameter is known, respectively. 
The experiment consists of generating 5000 samples of size $n = \{50, 100,150, \dots, 1000\}$,  with $\mathcal{G}_I^0(\alpha, \gamma, L)$ distribution. In this case $\alpha= -1.5$,  $\gamma = 1$, $L = 1$.
It can be observed that the proportion of estimates with error dramatically decreases as the sample size increases. 
This evidences that bias of the maximum likelihood estimates strongly depends on the sample size. 
Sample sizes greater than or equal to $750$ provide acceptable results but, in practical situations, one is often interested in smaller samples, e.g.\ for filters which compute estimates over windows of size $7\times 7$. 
The selected values of the parameters are arbitrary, in order to show an example of the maximum likelihood estimator behavior as the sample size increases.

As said, in each replication two independent samples were generated, and an estimate computed with each.
Each pair of estimates is then used to compute either $T_\alpha$ or $T_\gamma$, depending on the experiment.
Our main interest lies in the finite sample behavior of these test statistics.

\subsection{Finite Sample Size Behavior of $T_\alpha$}

For each sample size, we have five thousand samples of $T_\alpha$.
We will analyze the distribution of these test statistics, and the empirical size of the test when compared with the asymptotic result.

\begin{figure}[hbt]
    \centering
    \subfigure[Boxplots of $T_\alpha$ for $\gamma$ known.\label{fig:BoxPlotAlpha_gknown}]{\includegraphics[width=.49\linewidth]{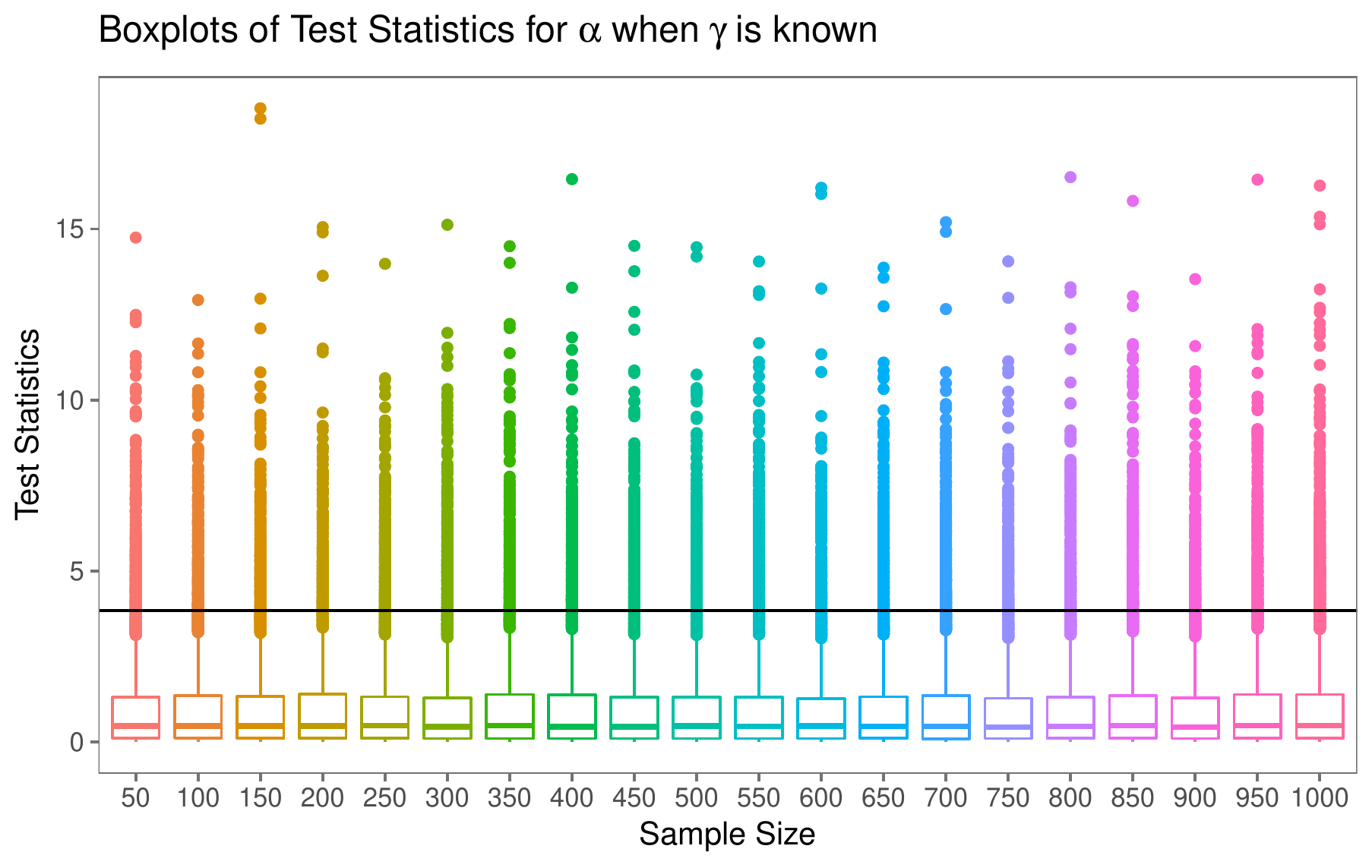}}
    \subfigure[Boxplots of $T_\gamma$ for $\alpha$ known.\label{fig:BoxPlotGamma_aknown}]{\includegraphics[width=.49\linewidth]{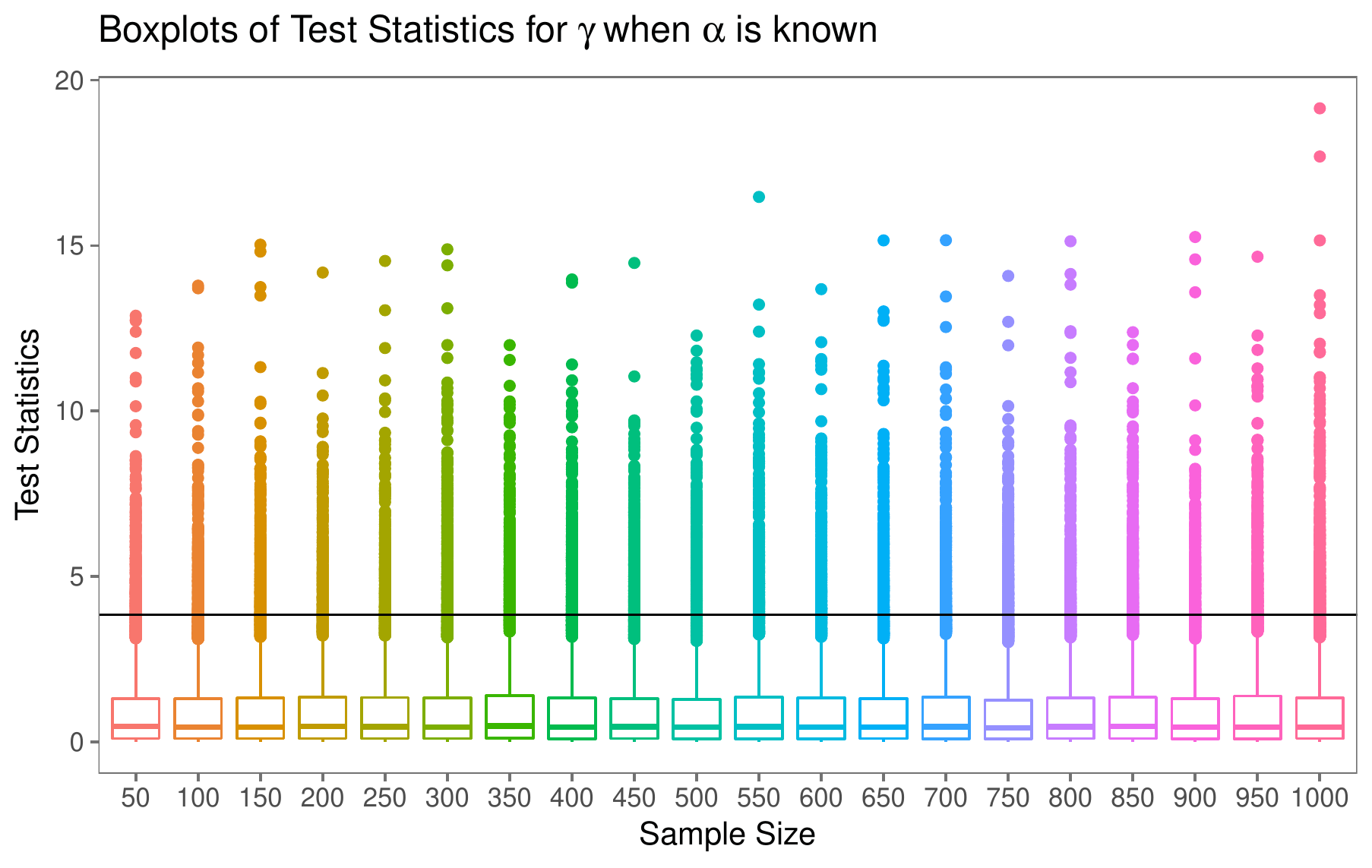}}
    \caption{Boxplots of tests statistics.}
    \label{fig:BoxPlotsTaTg}
\end{figure}

Fig.~\ref{fig:BoxPlotAlpha_gknown} shows the boxplots of the $T_\alpha$ test statistics for different sample sizes, along with the theoretical cut value at \SI{95}{\percent} (approximately \num{3.841459}, the \SI{0.95}{quantile} of the $\chi^2_1$ distribution).

Fig.~\ref{fig:DensitiesAlpha_gknown} shows the sample densities of the $T_\alpha$ test statistics for different sample sizes, along with the theoretical cut value at the \SI{95}{\percent} (approximately \num{3.841459}, the \SI{0.95}{quantile} of the $\chi^2_1$ distribution).

Neither Fig.~\ref{fig:BoxPlotAlpha_gknown} nor Fig.~\ref{fig:DensitiesAlpha_gknown} suggest any significant change of distribution of $T_\alpha$ when the sample size varies, an evidence that $n=50$ is a large enough sample size to attain the asymptotic properties.

Fig.~\ref{fig:pvaluesAlpha_gknown} presents the empirical size of $T_\alpha$ tests for different sample sizes, along with the theoretical cut value.
The minimal and maximal deviation between the empirical and theoretical $p$-values are, respectively, \SI{0.4}{\percent} and \SI{13.2}{\percent}.

\subsection{Finite Sample Size Behavior of $T_\gamma$}

For each sample size, we have five thousand samples of $T_\gamma$.
We will analyze the distribution of these test statistics, and the empirical size of the test with respect to the asymptotic value.

\begin{figure}[hbt]
    \centering
    \subfigure[Empirical densities of $T_\alpha$, for $\gamma$ known.\label{fig:DensitiesAlpha_gknown}]{\includegraphics[width=.49\linewidth]{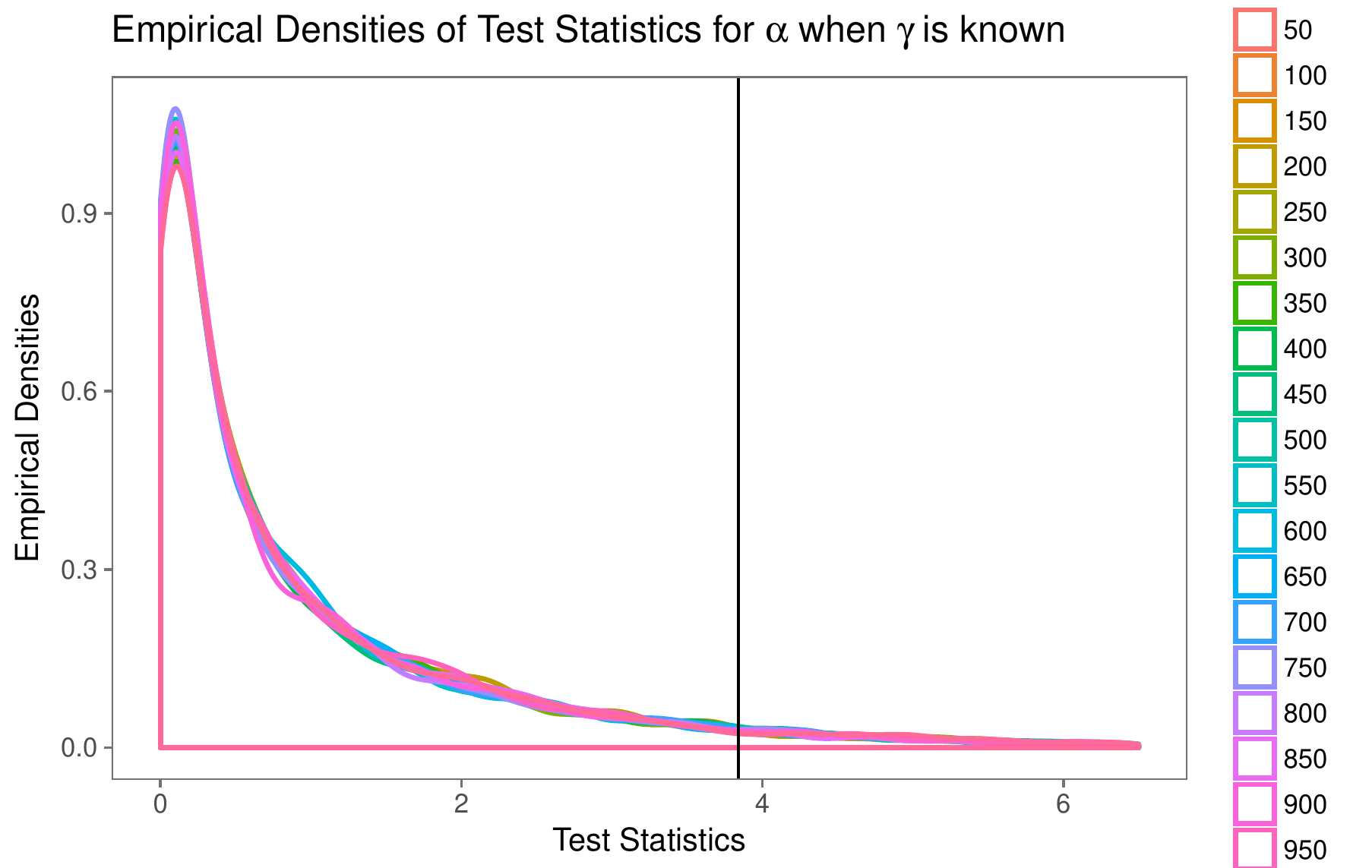}}
    \subfigure[Empirical densities of $T_\gamma$, for $\alpha$ known.\label{fig:DensitiesGamma_aknown}]{\includegraphics[width=.49\linewidth]{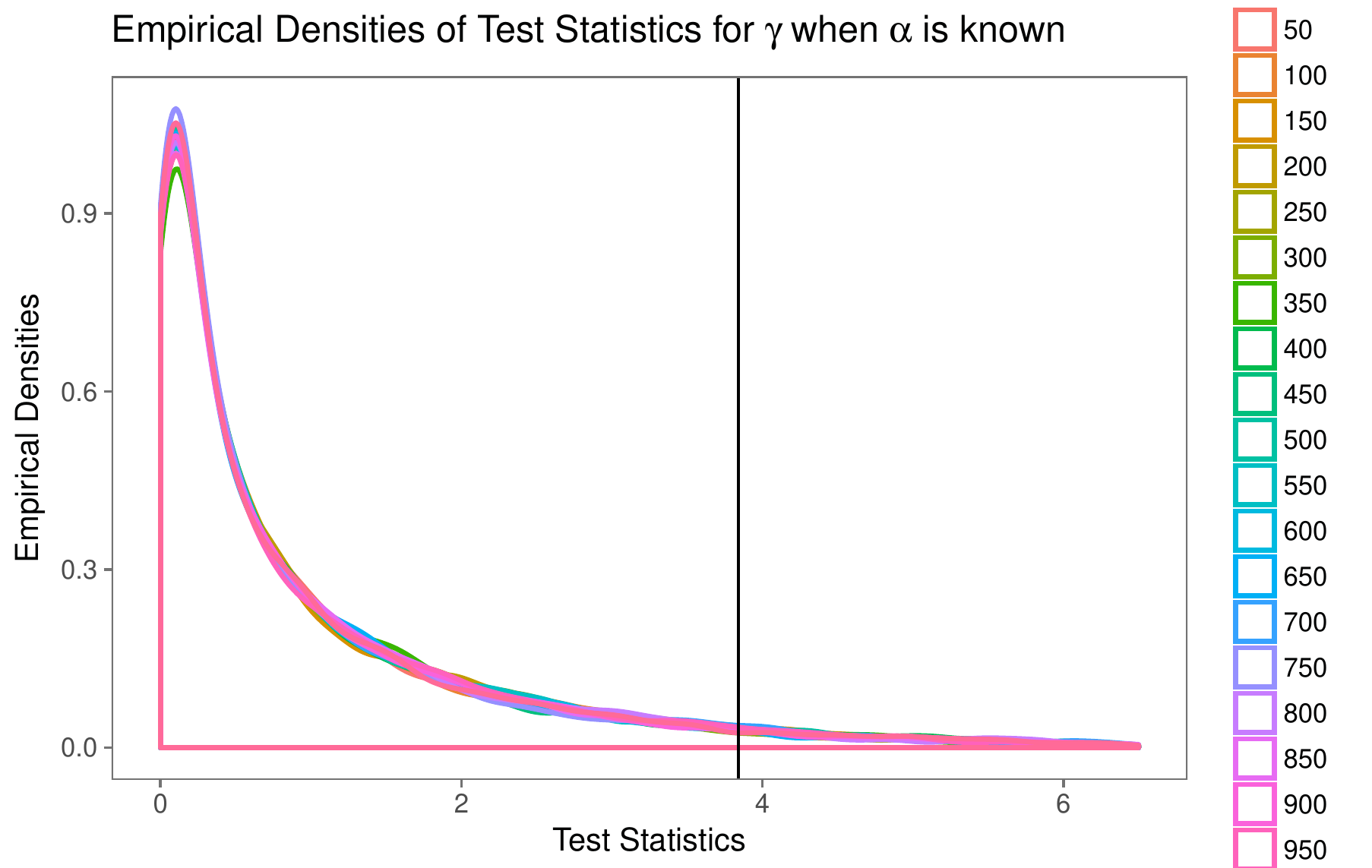}}
    \caption{Empirical densities of tests statistics.}
    \label{fig:EmpiricalDensitiesTaTg}
\end{figure}

Fig.~\ref{fig:BoxPlotGamma_aknown} shows the boxplots of the $T_\gamma$ test statistics for different sample sizes, along with the theoretical cut value at the \SI{95}{\percent} (approximately \num{3.841459}, the \SI{0.95}{quantile} of the $\chi^2_1$ distribution).
Fig.~\ref{fig:DensitiesGamma_aknown} shows the sample densities of the $T_\gamma$ test statistic, for different sample sizes,  along with the theoretical cut value at the \SI{95}{\percent} (approximately \num{3.841459}, the \SI{0.95}{quantile} of the $\chi^2_1$ distribution).

Neither Fig.~\ref{fig:BoxPlotGamma_aknown} nor Fig.~\ref{fig:DensitiesGamma_aknown} suggest any significant change of distribution of $T_\gamma$ when the sample size varies, evidence that $n=50$ is a large enough sample size to attain the asymptotic properties.
This motivates the use of a single model, namely the $\chi^2_1$ distribution, for computing quantiles.

\begin{figure}[hbt]
    \centering
    \subfigure[Empirical $p$-values of $T_\alpha$, $\gamma$ known.\label{fig:pvaluesAlpha_gknown}]{\includegraphics[width=.49\linewidth]{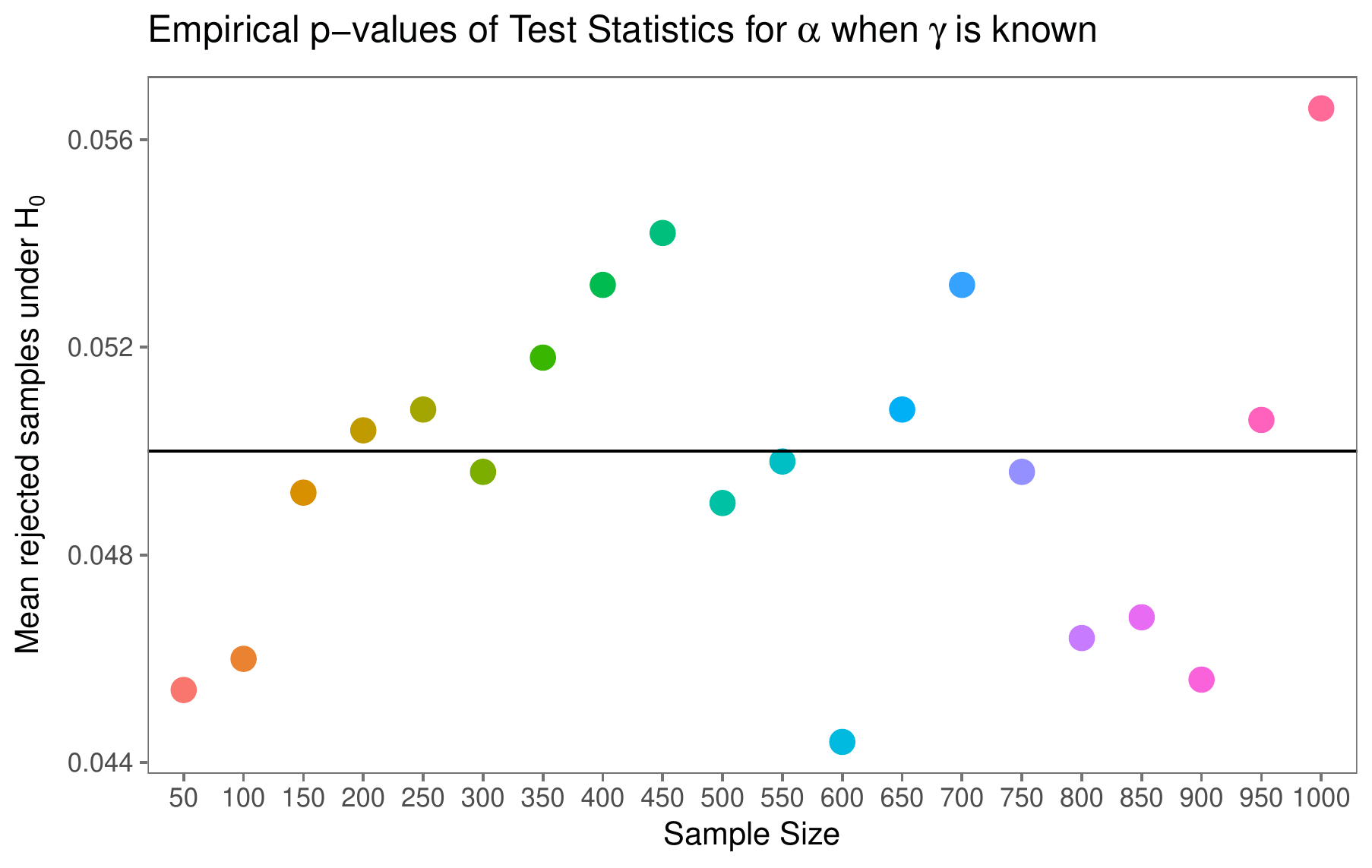}}
    \subfigure[Empirical $p$-values of $T_\gamma$, $\alpha$ known.\label{fig:pvaluesGamma_aknown}]{\includegraphics[width=.49\linewidth]{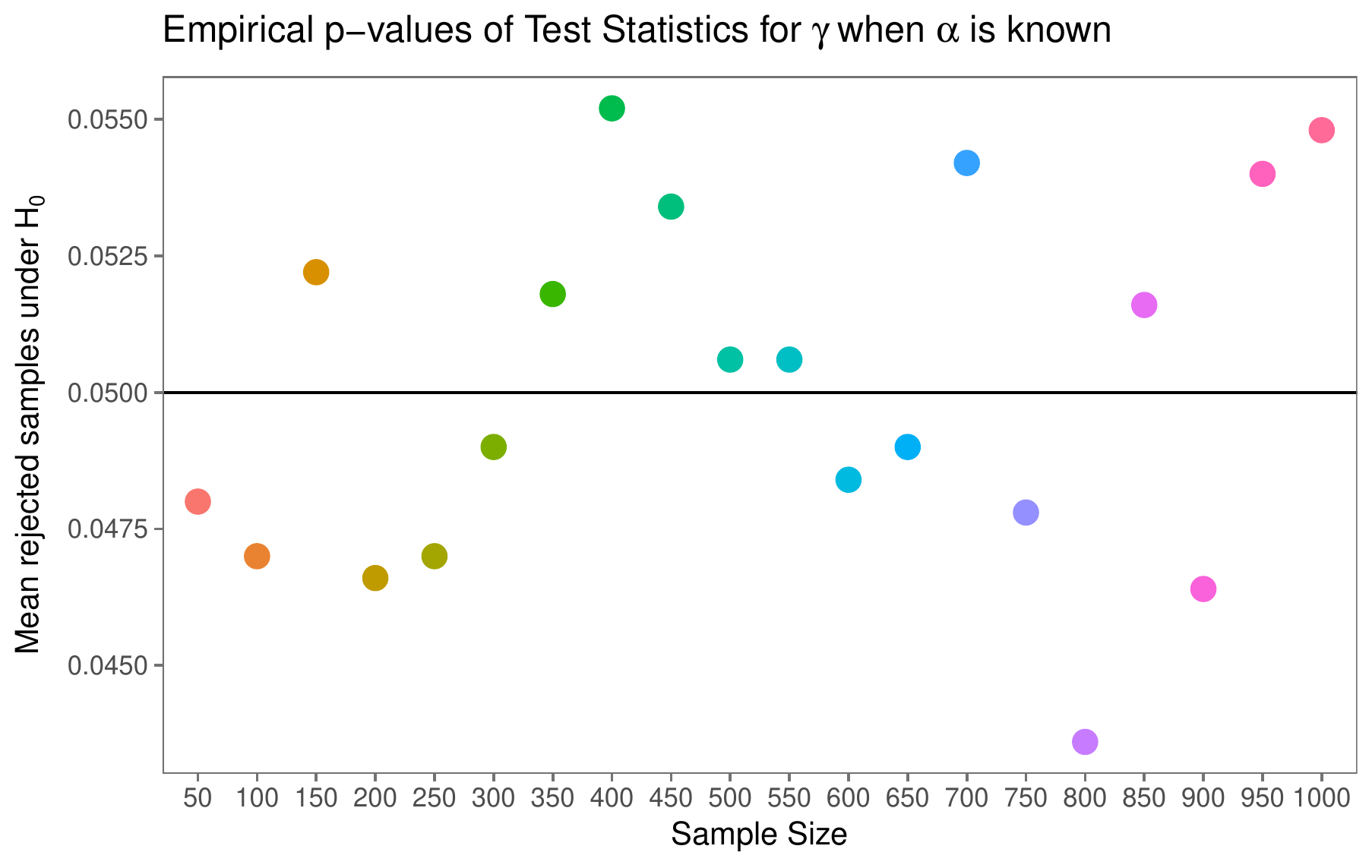}}
    \caption{Empirical size of tests statistics.}
    \label{fig:EmpiricalpvaluesTaTg}
\end{figure}

Figs.~\ref{fig:pvaluesGamma_aknown} presents the empirical size of $T_\gamma$ test for different sample sizes, along with the theoretical cut value.
The minimal and maximal deviation between the empirical and theoretical $p$-values are, respectively, \SI{1.2}{\percent} and \SI{12.8}{\percent}.

\section{Analysis of Two-parameter tests}\label{Sec:TwoParameter}

In this section, we analyze the more realistic situation of estimating both the scale and texture parameters, while assuming the number of looks known.
As mentioned, we opted for computing $(\widehat{\alpha}, \widehat{\gamma})$ the maximum likelihood estimator of  $({\alpha}, {\gamma})$ by maximizing the reduced log-likelihood function which, for $L$ known, is
\begin{equation}
\ell(\alpha,\gamma;L,\bm z) =  
n [\log\Gamma(L-\alpha) - \alpha\log\gamma - \log\Gamma(-\alpha)] +
(\alpha-L) \sum_{i=1}^n \log(\gamma+L z_i).
\label{eq:LogVerAlphaGamma}
\end{equation}
Again, the routine \texttt{maxLik} was the tool employed for maximizing~\eqref{eq:LogVerAlphaGamma}.

Whereas maximizing~\eqref{eq:LogVera_gknown} and~\eqref{eq:LogVerg_aknown} poses no numerical problem, \eqref{eq:LogVerAlphaGamma} has well-reported problems caused by cases where this likelihood becomes flat~\cite{FreryCribariSouza:JASP:04}.
In order to avoid such problems without introducing specialized techniques that depart from the concept of maximum likelihood, only solutions satisfying $(\widehat{\alpha}, \widehat{\gamma}) \in [15\alpha,0)\times (0, 15\gamma]$ where considered feasible.
The number of replications is computed over feasible solutions.

The parameter space of the study is the product of the sets
$\alpha \in \{-1.5, -3, -4\}$, 
$L \in \{1, 2\}$, and 
$n \in \{50(100)950, 5000\}$.
For each $\alpha$, the scale is $\gamma = -\alpha-1$, so the expected value is $1$.
Following the recommendations discussed in~\cite{busto92}, the number of replications changes with the sample size as $R=[R_{\max}/n]$;  we empirically found  $R_{\max}=5\times 10^6$  produces reliable results with an acceptable computational cost.

\begin{figure}[hbt]
    \centering
    \subfigure[Empirical densities of $\widehat{\alpha}$.\label{fig:AlphaBoth}]{\includegraphics[width=.49\linewidth]{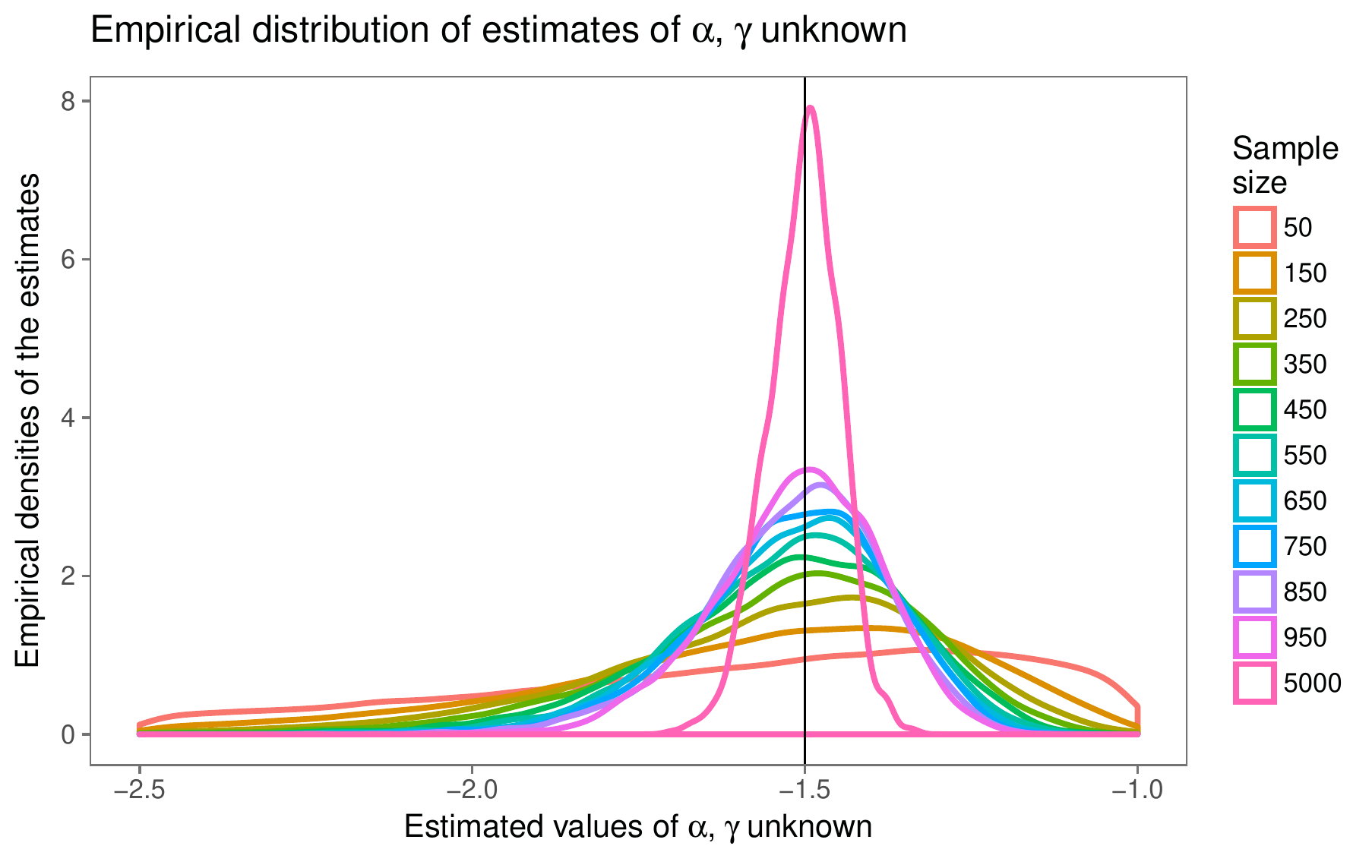}}
    \subfigure[Empirical densities of $\widehat{\gamma}$.\label{fig:GammaBoth}]{\includegraphics[width=.49\linewidth]{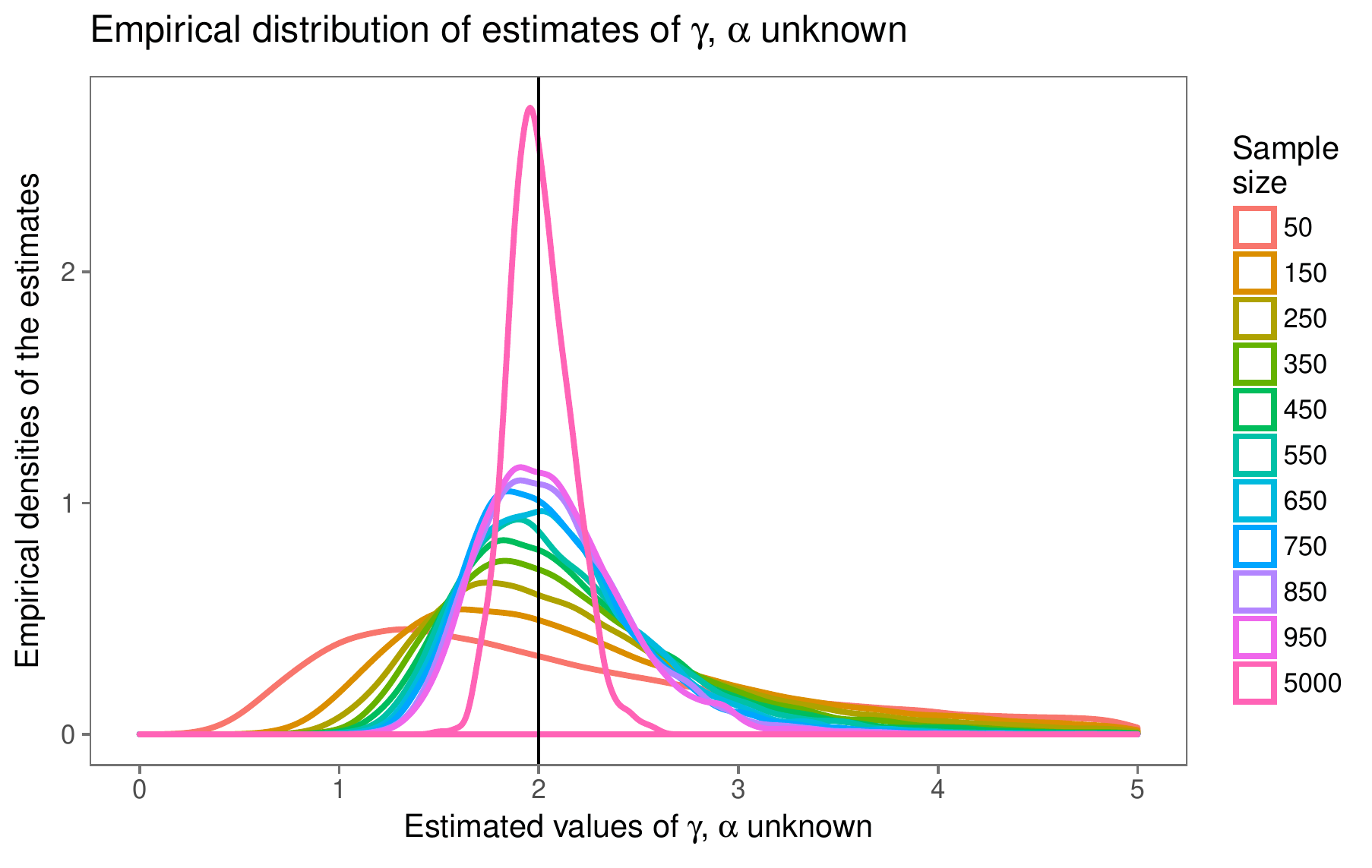}}
    \caption{Empirical densities of estimators when only $L=1$ is known, $\alpha=-3$ and $\gamma=2$.}
    \label{fig:DensitiesEstimatorsBothUnknown}
\end{figure}

The plots in Fig.~\ref{fig:DensitiesEstimatorsBothUnknown} show the empirical densities of the estimators of texture and scale, Fig.~\ref{fig:AlphaBoth} for the case $\alpha=-1.5$, $\gamma=0.5$ and $L=1$, Fig.~\ref{fig:GammaBoth} for the case $\alpha=-3$, $\gamma=2$ and $L=1$.

The difference between Figs.~\ref{fig:DensMLAlpha_gknown} and~\ref{fig:AlphaBoth} is noticeable in terms of spread and centrality.
The same observation holds when comparing figures~\ref{fig:DensMLGamma_aknown} and~\ref{fig:GammaBoth}.
The effect of missing the information of one parameter is, thus, remarkable.

%

\begin{figure}
    \centering
    \includegraphics[width=\linewidth]{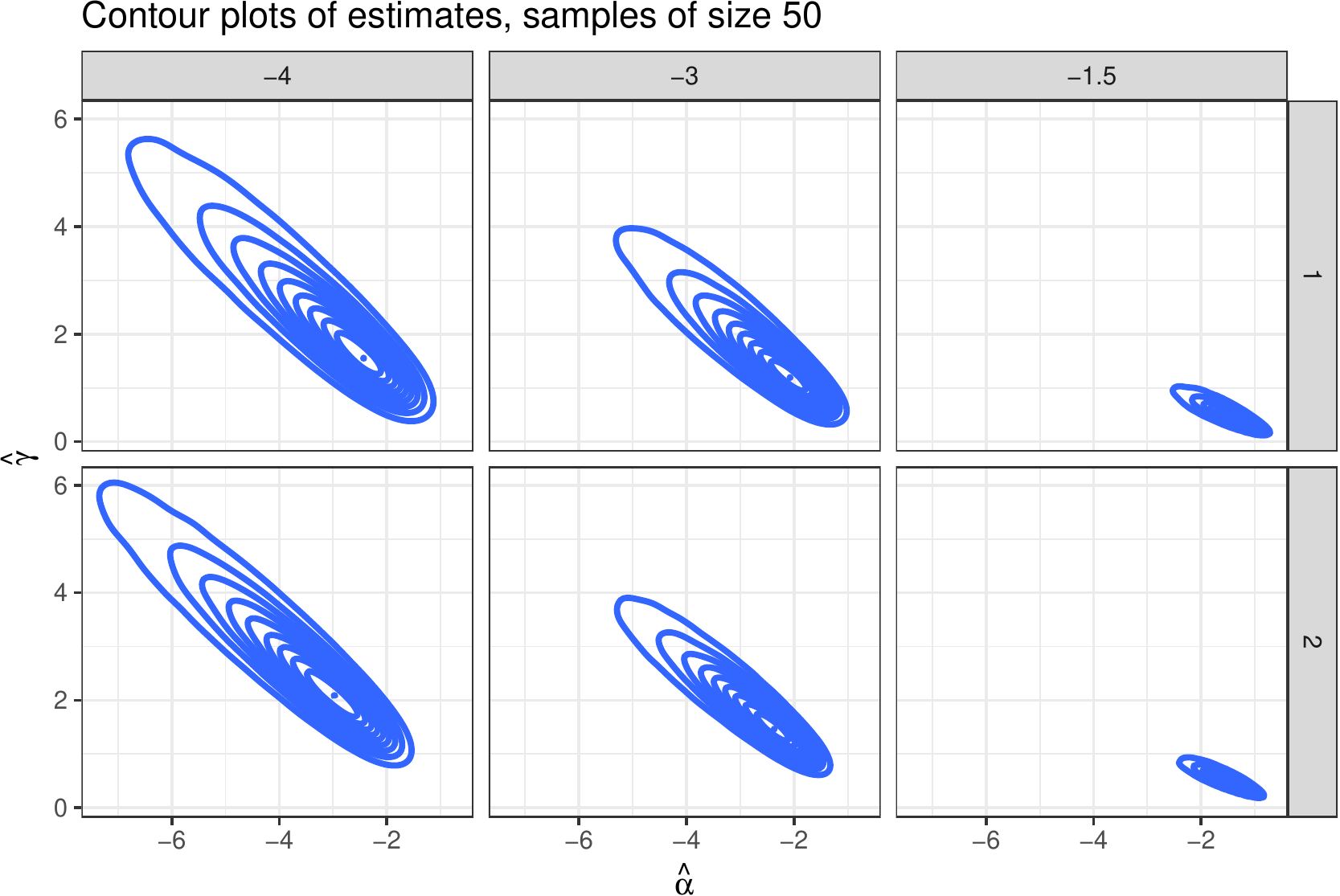}
    \caption{Contour plots of $(\widehat{\alpha},\widehat{\gamma})$ for all the cases considered and samples of size $n=50$.}
    \label{fig:ContourPlotsAlphaGamma}
\end{figure}

Fig.~\ref{fig:ContourPlotsAlphaGamma} shows the contour plots of the estimates $(\widehat{\alpha},\widehat{\gamma})$ for samples of size $n=50$ and all the cases here considered. 
This figure corroborates that it is not adequate to assume that $\widehat\alpha$ and $\widehat{\gamma}$ can be uncorrelated, let alone independent.

\begin{figure}
    \centering
    \includegraphics[width=\linewidth]{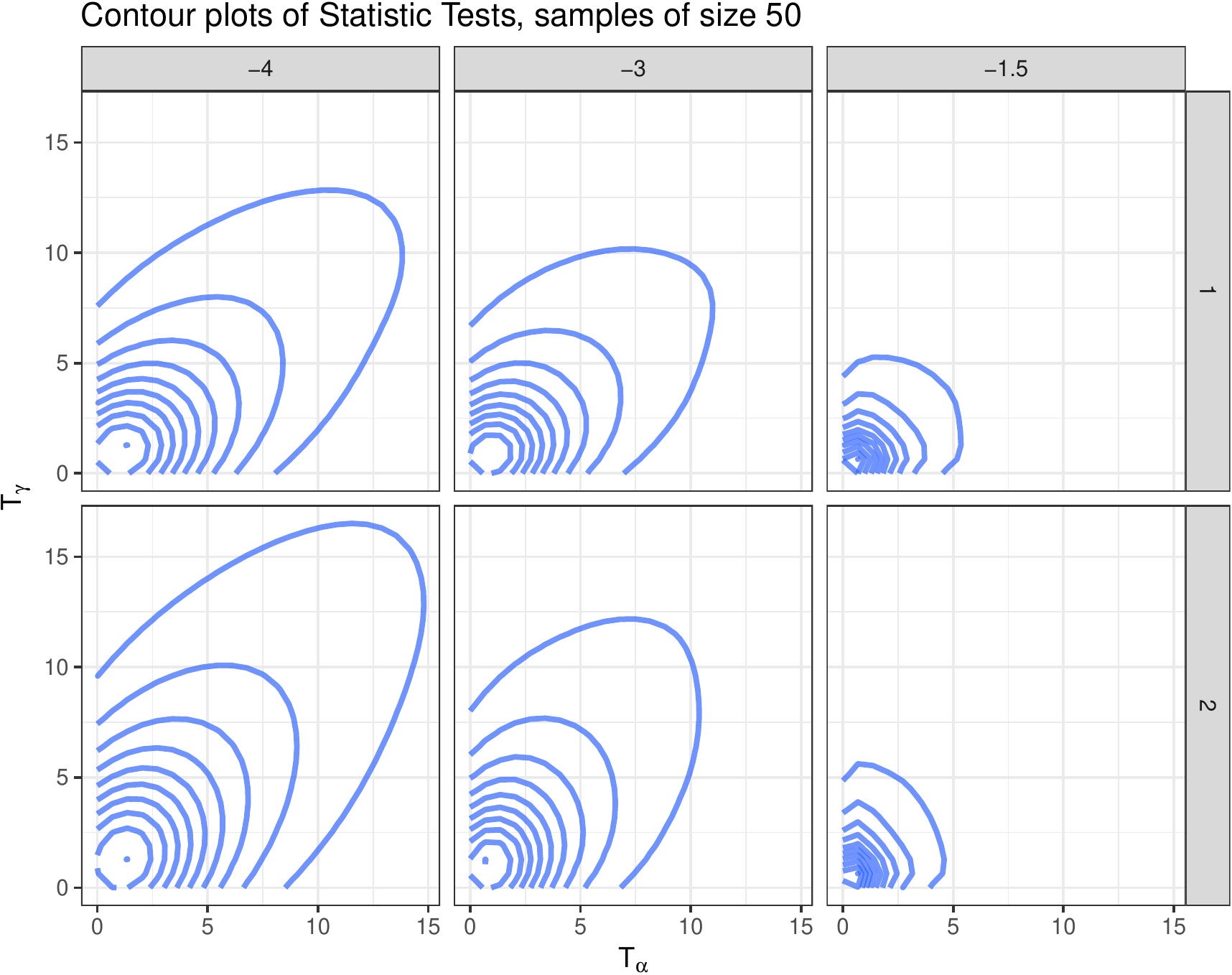}
    \caption{Contour plots of $(T_\alpha,T_\gamma)$ for all the cases considered and samples of size $n=50$.}
    \label{fig:ContourPlotsTAlphaTGamma}
\end{figure}

This relationship between estimators is also exhibited by the tests statistics that use them.
Fig.~\ref{fig:ContourPlotsTAlphaTGamma} shows the contour plots of $(T_\alpha,T_\gamma)$.

In practice, one needs to discriminate regions with unknown texture and scale, so a test statistic for both parameters, say $T_{\alpha,\gamma}$ is required.

The strong relationship between $\widehat{\alpha}$ and $\widehat{\gamma}$ is evident, so is the same relationship between test statistics, therefore just adding $T_\alpha$ and $T_\gamma$ and assuming that the sum follows a $\chi^2_2$ law might not be a good idea.
This justifies the following analysis which aims at finding relevant properties of two-to-one transformations of $(T_\alpha,T_\gamma)\to T_{\alpha,\gamma}$, in search for a test statistic for assessing the null hypothesis of having two samples from the same $\mathcal{G}^0$ distribution.
We analyze the following test statistics:
\begin{align}
T^1_{\alpha,\gamma} &=  \sqrt{T^2_{\alpha=(\widehat{\alpha}_1+\widehat{\alpha}_2)/2} + T_\gamma^2}, \label{eq:SqrtSumSquares}\\
T^2_{\alpha,\gamma} &=  \frac{T_{\alpha=(\widehat{\alpha}_1+\widehat{\alpha}_2)/2} + T_\gamma}{2}, \label{eq:MeanTests}\\
T^3_{\alpha,\gamma} &=  \max\Big\{\frac{T_{\alpha=(\widehat{\alpha}_1+\widehat{\alpha}_2)/2}}{T_\gamma}, \frac{T_\gamma}{T_{\alpha=(\widehat{\alpha}_1+\widehat{\alpha}_2)/2}} \Big\} \label{eq:MaxRatio}
\end{align}
Eqs.~\eqref{eq:SqrtSumSquares}, \eqref{eq:MeanTests} and~\eqref{eq:MaxRatio} are combinations of statistics for a single free parameter, given in Eqs.~\eqref{eq:Ta} and~\eqref{eq:Tg},  but their distributions are unknown and, thus, we can not apply a hypothesis test to decide if two samples come from the same distribution or not. 
So, to solve this problem, we estimate these distributions using permutation methods, as explained in Section~\ref{permutationMethods}.

\subsection{Permutation Methods}
\label{permutationMethods}

Permutation methods are a type of statistical significance test which can be applied to statistics with unknown distribution. 
They were developed by R.\ Fisher and E.\ J.\ G.\ Pitman~\cite{Fisher1934}.
The authors of Refs.~\cite{PermutationMethods2011,WICS:WICS1429}  explain the advantages of this type of tests. 
There are at least two kinds of permutation tests:
\begin{description}[\IEEEsetlabelwidth{\textbf{Random}}]
    \item[\textbf{Exact:}] in which all possible reorganizations of the sample are considered. This kind has high computational cost, depending on the sample size.   
    \item[\textbf{Random:}] which consider a certain amount of permutations, usually $1000$ or $10000$. They are more appropriate if the sample size is large.
\end{description}

In this work we test if two samples $X \sim \mathcal{G}^0(\alpha_1, \gamma_1, L)$  and  $ Y\sim \mathcal{G}^0(\alpha_2, \gamma_2, L)$ are from the same distribution, then we pose the null hypothesis $H_0: (\alpha_1, \gamma_1) = (\alpha_2, \gamma_2)$ and we want to know the probability of rejecting it.
With this objective, 
we estimate the empirical distributions of the tests $T^i_{\alpha,\gamma}, \; i = 1,2,3$ from equations~\eqref{eq:SqrtSumSquares}, \eqref{eq:MeanTests} and~\eqref{eq:MaxRatio} by means of the following steps. 
For more information see~\cite{PermutationTestandStatisticalSignificance}.
\begin{enumerate}
    \item Choose a statistic $T^i_{\alpha,\gamma}$, $i=1,2,3$ from  Eqs.~\eqref{eq:SqrtSumSquares}, \eqref{eq:MeanTests} or~\eqref{eq:MaxRatio}.
    
    \item Generate $\bm z_1$ and $\bm z_2$ random samples of sizes $m$ and $n$, respectively, both from the same $\mathcal{G}^0(\alpha,\gamma, L)$ distribution granting the null hypothesis. Let  $\text{perm}$  be the number of permutations; in our experiment $\text{perm} = 1000$. 
    
    \item Compute the estimates $(\widehat{\alpha}_1, \widehat{\gamma}_1)$ and  $(\widehat{\alpha}_2, \widehat{\gamma}_2)$ with each sample. 
    
    \item\label{paso3} Calculate the observed statistic value, $T^i_{\alpha,\gamma}$, with the data from $\bm z_1$ and $\bm z_2$.
    
    \item Repeat for $k = 1,\dots, \text{perm}$:
    \begin{itemize}
        \item\label{paso4} Shuffle de joint sample $\bm z = (\bm z_1,\bm z_2)$ and divide it in two groups of sizes $m$ and $n$, say $\bm z_1^k$ and $\bm z_2^k$, respectively.
        
        \item\label{paso5} Compute the estimates $(\widehat{\alpha}_1, \widehat{\gamma}_1)$ and  $(\widehat{\alpha}_2, \widehat{\gamma}_2)$ for each sample $\bm z_1^k$ and $\bm z_2^k$. 
        
        \item \label{paso6} Calculate the  statistic value using the permuted samples, $T^p_{\alpha,\gamma}(k)$.
        
        \item Compare the observed statistic value calculated in Step~\ref{paso3} with the statistic computed after permutation $T_{\alpha, \gamma}^p(k)$.
    \end{itemize}

  \item The proportion of
    differences equal to or larger than the observed statistic value
    serve as the $p\text{-value}$ for the permutation test, or:
    \begin{equation}
    p\text{-value} =
    \frac{\#\{k : T_{\alpha, \gamma}^p(k) \geq T^i_{\alpha, \gamma} \}}{\text{perm}}.
    \end{equation}

    \item If $p\text{-value}< \eta$, the null hypothesis is rejected at level $\eta$. 

\end{enumerate}

\subsection{Results of applying Permutation Methods}

We use devised Monte Carlo experiments to quantify empirical rejection rate (R-rate) generated by the proposed tests, under the Null Hypothesis. 
The experiment is repeated $500$ times.

Table~\ref{TablaDeFalsosNegativos} shows the results of applying the permutation test to the statistic given in Eqs.~\eqref{eq:SqrtSumSquares}, \eqref{eq:MeanTests} and~\eqref{eq:MaxRatio}, for values of $L = \{1,2\}$, $\alpha = \{-1.5,-4\}$ and $\gamma = -\alpha- 1$, at level $\eta =0.05$. 
For lack of space, we present only the results for $n = 50$, $n = 550$ and $n = 5000$, corresponding to small, medium and large samples.
We inform the rejection rate under the null hypothesis (false negative rate) which is the estimated test size. 
Tests $T^1$ and $T^2 $ exhibit the closest empirical sizes to the nominal level.  
It can be observed that if the sample size increases, the false negative rate is not necessarily reduced.

\begin{table}[hbt]
    \centering
    \caption{Rejection Rates for the proposed statistics under the null hypothesis.}
    \label{TablaDeFalsosNegativos}
    \begin{tabular}{@{}crrrrr@{}}
        \toprule
        $L$& $\alpha$   & $n$  & R-rate $T^1$  & R-rate $T^2$& R-rate $T^3$\\ \midrule
        \multirow{3}{*}{1} & \multirow{3}{*}{$-1.5$} & $50$ & $0.048$  & $0.058$  & $0.075$  \\
        \multirow{3}{*}{} & \multirow{3}{*}{} & $550$  & $ 0.056$ & $0.050$  & $0.051$ \\
        \multirow{3}{*}{} & \multirow{3}{*}{} & $5000$  & $0.044$ & $0.048$  & $0.058$ \\
        \hline
        \multirow{3}{*}{1} & \multirow{3}{*}{$-4$} & $50$  & $0.046$ & $0.046$  & $0.049$  \\
        \multirow{3}{*}{} & \multirow{3}{*}{} & $550$  & $0.056$ & $0.056$  & $0.045$  \\
        \multirow{3}{*}{} & \multirow{3}{*}{} & $5000$  & $0.046$ & $0.046$  & $0.050$  \\
        \midrule        
        \multirow{3}{*}{2} & \multirow{3}{*}{-1.5} & $50$  &  $0.060$ & $0.056$  & $0.05$ \\
        \multirow{3}{*}{} & \multirow{3}{*}{} & $550$ & $0.052$ & $0.052$ & $0.043$  \\        
        \multirow{3}{*}{} & \multirow{3}{*}{} & $5000$  & $0.052$ & $0.042$  & $0.059$ \\
        \hline
        \multirow{3}{*}{2} & \multirow{3}{*}{-4} & $50$  & $0.06$ & $0.060$  & $0.051$ \\
        \multirow{3}{*}{} & \multirow{3}{*}{} & $550$  & $0.038$ & $0.038$  & $0.035$ \\
        \multirow{3}{*}{} & \multirow{3}{*}{} & $5000$  & $0.048$ & $0.048$  & $0.055$  \\
        \bottomrule
    \end{tabular}
\end{table}

\begin{figure}[hbt]
    \centering
\includegraphics[width=.95\linewidth]{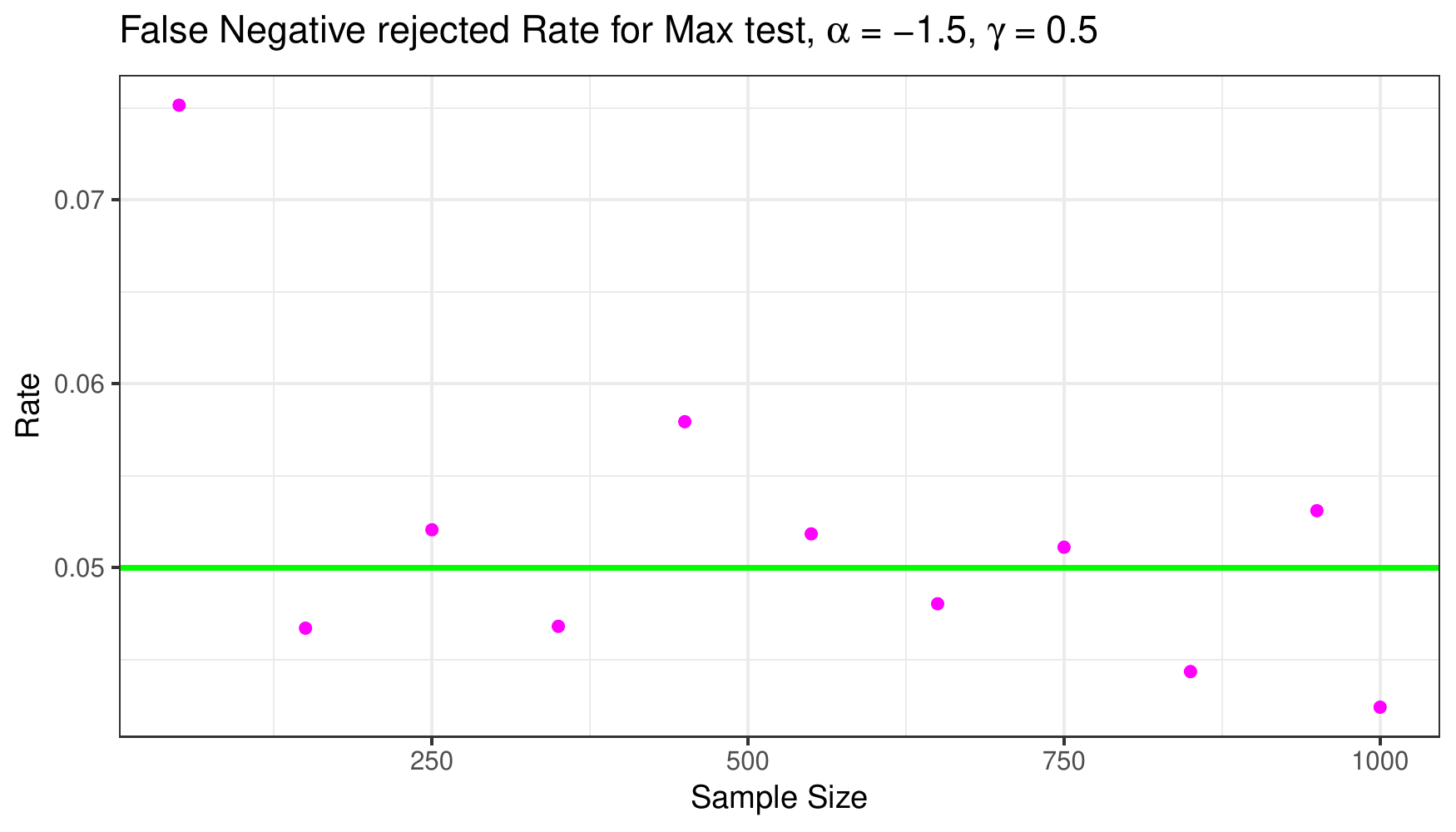}
\caption{False Negative Rate depending on the sample size, under the Null Hypothesis.}
    \label{fig:FalsePositiveRate}
\end{figure}

Figure~\ref{fig:FalsePositiveRate} shows the false negative rate for the test in Eqs.~\ref{eq:MaxRatio}, under the Null Hypothesis depending on the sample size, for $\alpha = -1.5$, $\gamma = 0.5$, $L =1$. It can be observed that the false negative rate fluctuates around the value of the level $\eta = 0.05$, represented with a green straight line and the highest value of the false negative rate is given for the sample size $n = 50$.

\subsection{Application in Edge Detection}

In this section, we present an application of the proposed method to the problem of edge detection in actual SAR images. 
Gambini et al.~\cite{GambiniSC08} proposed a general and flexible algorithm for edge detection which is based on finding, in a narrow strip of data, the point where there is maximum evidence of a change of properties.
Naranjo-Torres et al.~\cite{GeodesicDistanceGI0JSTARS} used a geodesic distance between models as a measure of this change, assuming the $\mathcal{G}_I^0$ distribution with known scale parameter. 
In this work, we use the same algorithm but considering two parameters unknown: texture $\alpha$, and scale $\gamma$. 
In order that this work is self-contained,  we briefly explain the algorithm. 
For more information see~\cite{GeodesicDistanceGI0JSTARS}.

Let $I$ be an actual SAR image of $m$ lines and $n$ columns of pixels. 
In this application, we use only one line of data, i.e., a strip of size $1\times n$.
In each step $3\leq k\leq n-3$, we divide the line in two disjoint samples, $S_1(k)=(z_1,\dots,z_k)$ and $S_2(k)=(z_{k+1},\dots,z_{n})$ used to estimate the parameters $(\widehat\alpha_1, \widehat\gamma_1)(k)$ and $(\widehat\alpha_2, \widehat\gamma_2)(k)$, respectively, by maximum likelihood.
Then, the $p$-value  $p(k)$ is computed using the method described in Section~\ref{permutationMethods}. 

Finally, we estimate the transition point as the position at which $p(k)$ is minimum: $ \widehat{col} = \arg\min_{k}p(k)$.
The method is sketched in Algorithm~\ref{Algoritmo}, where $I$ is the original image,  $m$ and $n$ are the numbers of rows and columns of the input image.
Notice that the minimum sample size is set to three observations.

\begin{algorithm}[hbt]
\caption{Edge Detection by the geodesic distance of the $\mathcal{G}_I^0$ distribution with two unknown parameters.}\label{Algoritmo}
    \begin{algorithmic}[1]
        \STATE input: $I$, $m$, $n$
        \FOR {each line of $I$, $i = 1,\dots,m$} 
        \FOR{$k = 3, \dots, n-3$}
        \STATE Divide the line in two samples $S_1(k)=(z_1,\dots,z_k)$ and $S_2(k)=(z_{k+1},\dots,z_{n})$.
        \STATE Estimate $(\alpha,\gamma)$ by maximum likelihood in each sample, obtaining $(\widehat{\alpha}_1,\widehat{\gamma}_1) (k)$ and $(\widehat{\alpha}_2 ,\widehat{\gamma}_2 ) (k)$.
        \STATE Compute $T(k) = T_{\widehat{\alpha}(k),\widehat{\gamma}(k)}$ using Eqs.~\eqref{eq:SqrtSumSquares}, \eqref{eq:MeanTests} or~\eqref{eq:MaxRatio}.
        \STATE Consider the array of statistics between the  pairs of samples:
        $\bm T=\{T(k), \; 3\leq k\leq m-3\}$ and compute the array of $p$-values $ \bm P = \{p(k), \; 3\leq k\leq m-3\}$.
        \STATE Find the column where the array $\bm P$ is minimized, which corresponds to the transition point on the line $i$: 
        \begin{equation*}
            \widehat{col} = \arg \min_{k} \bm p(k),
        \end{equation*}
        \ENDFOR
        \ENDFOR
    \end{algorithmic}

\end{algorithm}

\begin{figure}[hbt]
    \centering
    \subfigure[SAR image and the region used.\label{fig:OriginalImage}]{\includegraphics[bb=100 100 250 250, clip=true, width=.5\linewidth]{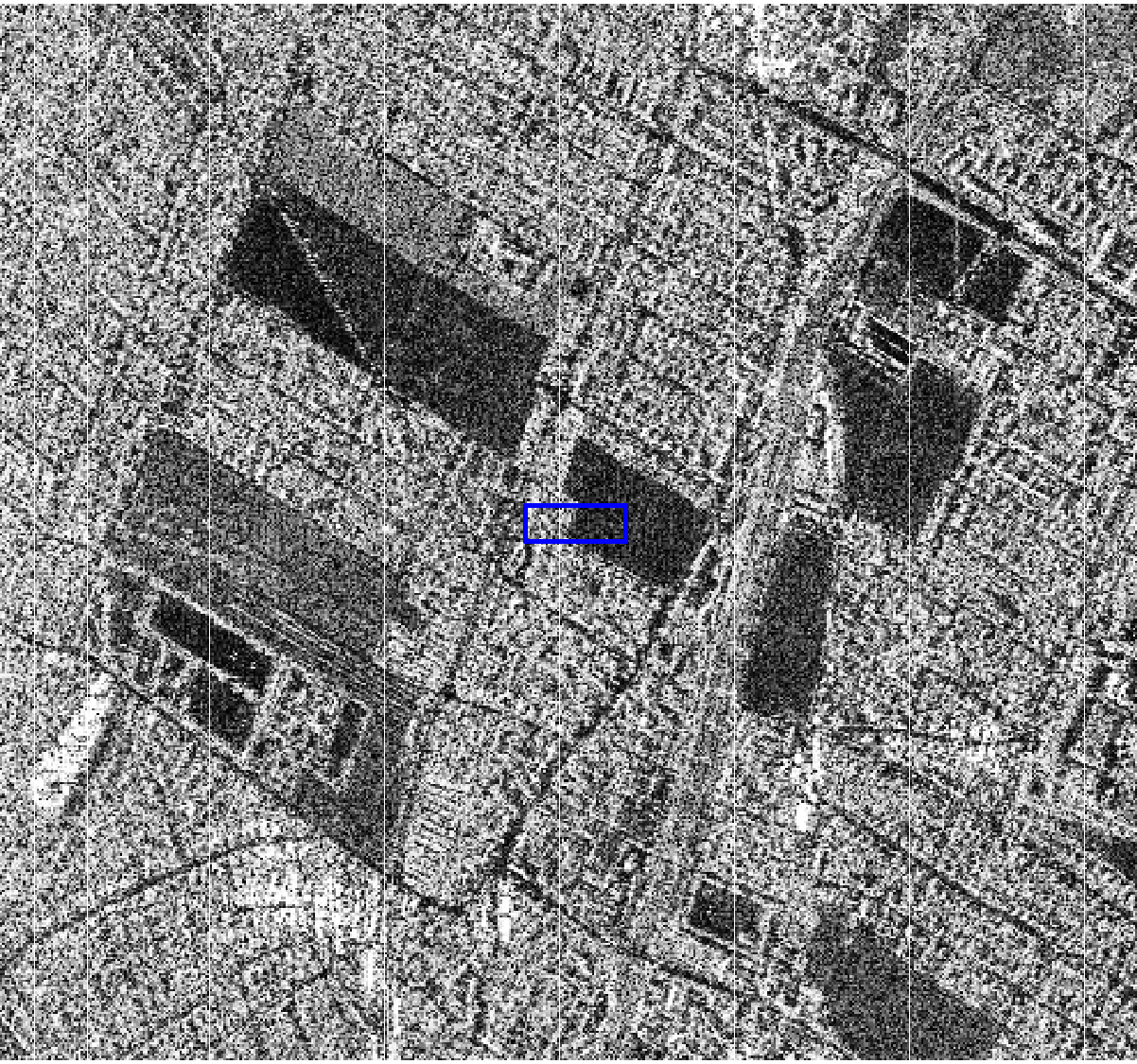}}

    \subfigure[Edge points found over each line of the region. \label{fig:BorderDetection}]{\includegraphics[width=.7\linewidth]{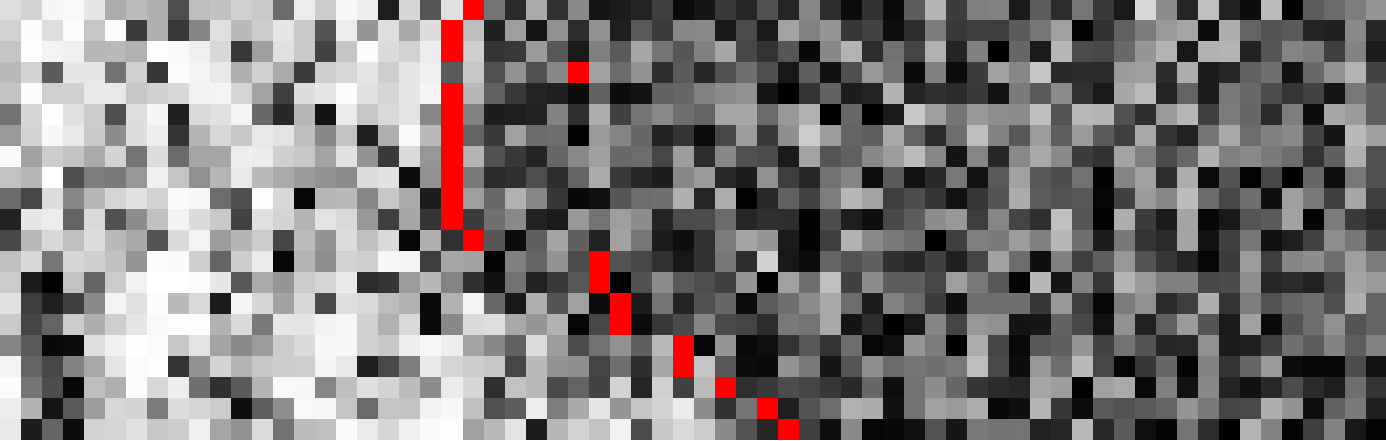}}
    \caption{Results of applying the edge detector to actual data using $T^1_{\alpha,\gamma}$. }
    \label{fig:EdgeDetectors}
\end{figure}

Figure~\ref{fig:EdgeDetectors} shows the results of applying the edge detector algorithm. Figure~\ref{fig:OriginalImage} shows the SAR image, and presents the area where the edge detection was performed. 
Figure~\ref{fig:BorderDetection} shows the result of applying the edge detector to each line in a selected region.

\section{Conclusions and Future Work}
\label{conclu}
Unable to calculate the geodesic distance of the $\mathcal{G}^0$ distribution depending on two free parameters, we carried out a study dedicated to evaluating the possibility of using a combination of tests based on the geodesic distance with a single unknown parameter, as calculated in Ref.~\cite{GeodesicDistanceGI0JSTARS}.

We compare three statistics whose distributions are unknown. 
We use permutation methods to estimate their empirical distributions. 

The results show that, under the null hypothesis, the false negative rate fluctuates around the rejection level, even with small samples. 
It can be observed that if the sample size increases, the false negative rate is not necessarily reduced; this encourages us to continue the investigations with small samples. 
The results are promising and can be readily employed in speckled image processing and analysis.

\appendix
Simulations were performed using the \texttt{R} language and environment for statistical computing version~3.0.2~\cite{Rmanual}.

\end{document}